\documentclass[floats,floatfix,showpacs,amssymb,prd,superscriptaddress,nofootinbib,twocolumn,aps,eqsecnum]{revtex4-1}

\usepackage{amsmath, amsthm, amsfonts, amssymb, latexsym}
\usepackage{graphicx}
\usepackage{color}
\usepackage{epsfig}
\usepackage{url}

\def\be{\begin{equation}}
\def\ee{\end{equation}}
\def\beq{\begin{eqnarray}}
\def\eeq{\end{eqnarray}}

\usepackage[unicode]{hyperref}
\hypersetup{
    unicode=true,          
    pdftitle={Collisions of charged black holes},
    pdfsubject={Numerical Relativity},
    pdfcreator={LaTeX},          
    pdfproducer={LaTeX},         
    pdfkeywords={General Relativity, Numerical Relativity, Black Holes},
    colorlinks=false,            
  }

\begin{document}

\title{Collisions of charged black holes}

  \author{Miguel~Zilh\~ao}\email{mzilhao@fc.up.pt}
  \affiliation{
    Departamento de F\'\i sica da Universidade de Aveiro and I3N, 
    Campus de Santiago, 3810-183 Aveiro, Portugal
  }
  \affiliation{
    Perimeter Institute for Theoretical Physics, Waterloo, Ontario N2L 2Y5, Canada
  }
  \affiliation{
    Centro de F\'\i sica do Porto, Departamento de F\'\i sica e Astronomia, 
    Faculdade de Ci\^encias da Universidade do Porto, Rua do Campo Alegre, 4169-007 Porto, Portugal
  }

  \author{Vitor~Cardoso}
    \affiliation{
    Centro Multidisciplinar de Astrof\'\i sica --- CENTRA,
    Departamento de F\'\i sica, Instituto Superior T\'ecnico --- IST,
    Av. Rovisco Pais 1, 1049-001 Lisboa, Portugal 
  }
  \affiliation{
    Department of Physics and Astronomy, The University of Mississippi,
    University, MS 38677-1848, USA
  }

 \author{Carlos~Herdeiro}
    \affiliation{
    Departamento de F\'\i sica da Universidade de Aveiro and I3N, 
    Campus de Santiago, 3810-183 Aveiro, Portugal
  }
  
   \author{Luis~Lehner}
    \affiliation{
    Perimeter Institute for Theoretical Physics, Waterloo, Ontario N2L 2Y5, Canada
  }
   \affiliation{
    Department of Physics, University of Guelph, Guelph, Ontario N1G 2W1, Canada
  }

  \author{Ulrich~Sperhake}
    \affiliation{
    Institut de Ci\`encies de l'Espai (CSIC-IEEC), Facultat de Ci\`encies, 
    Campus UAB, E-08193 Bellaterra, Spain
  }
    \affiliation{
    California Institute of Technology,
    Pasadena, CA 91125, USA
  }
    \affiliation{
    Centro Multidisciplinar de Astrof\'\i sica --- CENTRA,
    Departamento de F\'\i sica, Instituto Superior T\'ecnico --- IST,
    Av. Rovisco Pais 1, 1049-001 Lisboa, Portugal
  }

\begin{abstract}
We perform fully non-linear numerical simulations of charged-black-hole
collisions, described by the Einstein-Maxwell equations, and contrast
the results against analytic expectations. We focus on head-on collisions
of non-spinning black holes, starting from rest and with the same charge
to mass ratio, $Q/M$.
The addition of charge to black holes introduces a new interesting
channel of radiation and dynamics, most of which seem to be captured by
Newtonian dynamics and flat-space intuition.  The waveforms can be
qualitatively described in terms of three stages;
(i) an infall phase prior to the formation of a common
apparent horizon; (ii) a nonlinear merger phase
which corresponds to a peak in gravitational and electromagnetic
energy; (iii) the ringdown marked by an oscillatory pattern with
exponentially decaying amplitude and
characteristic frequencies that are in good agreement with perturbative
predictions.
We observe that the amount of gravitational-wave energy generated throughout
the collision
decreases by about three orders of magnitude
as the charge-to-mass ratio
$Q/M$ is increased from $0$ to $0.98$. We interpret this decrease
as a consequence of the smaller
accelerations present for larger values of the charge.
In contrast, the {\em ratio} of energy carried by
electromagnetic to gravitational radiation increases, reaching about
$22\%$ for the
maximum $Q/M$ ratio explored, which is in good agreement with
analytic predictions.
\end{abstract}


\maketitle


\section{Introduction}
In recent years, numerical relativity (NR) has generated a wealth of
information about astrophysical black-hole-binary systems;
see~\cite{Pretorius:2005gq,Baker:2005vv,Campanelli:2005dd}
for the first complete simulations and e.g.~\cite{Pretorius:2007nq,Palenzuela:2009yr,Palenzuela:2010nf,
Centrella:2010mx,Campanelli:2010ac,Sperhake:2011xk,Pfeiffer:2012pc}
for a representative list of more recent studies.
Results about the dynamics of black holes thus obtained are now
actively employed in techniques and searches for gravitational
wave signals in present and future generation gravitational wave detectors~\cite{Baumgarte:2006en,Aylott:2009tn,Abadie:2011kd,Ajith:2012tt,Virgo:2012aa}.

While black-hole binaries interacting with
electromagnetic fields and plasmas have been the subject of recent
numerical studies
(e.g.~\cite{Palenzuela:2009yr,Palenzuela:2010nf}), the dynamics
of binary systems of charged, i.e.~Reissner-Nordstr\"om (RN), black holes
remain unexplored territory. Perhaps, this is due to the expectation
that astrophysical black holes carry zero or very small charge; in particular, black holes with mass $M$, charge $Q$ and angular momentum
$aM^2$ are expected to discharge very quickly if
$Q/M \gtrsim 10^{-13}(a/M)^{-1/2}(M/M_{\odot})^{1/2}$~\cite{Gibbons:1975kk,Blandford:1977ds}.

In spite of this expectation,
however, there is a good deal of
motivation for detailed investigations of the dynamics of charged black holes.

We first note that
RN black holes possess a unique property amongst the black hole
solutions of Einstein-Maxwell theory in four dimensions. They
possess an extremal limit which can be used to construct a static, regular
(on and outside the event horizon) multi-black hole configuration~\cite{Hartle:1972ya} (described by the Majumdar-Papapetrou solution~\cite{Majumdar:1947eu,Papapetrou:1948jw}).  This configuration can be
interpreted as an exact cancellation, at each point, of the attractive
(gravitational) and repulsive (electromagnetic) interactions---a
\textit{no force condition}.  This condition is typically (but not
always) associated to supersymmetric configurations and indeed the
extremal RN solution is the only black hole solution in four dimensional
Einstein-Maxwell theory that admits Killing spinors, when the theory
is regarded as the bosonic sector of $\mathcal{N}=2$ Supergravity~\cite{Gibbons:1982fy,Tod:1983pm}. A natural question concerning the
modelling of RN black holes in NR is how close can we get to extremality
and hence consider the dynamics of these very special
black holes.  The ability to model such systems could provide interesting
applications. For instance it is
possible to study analytically the dynamics of a perturbed
Majumdar-Papapetrou solution in the so-called
moduli space approximation~\cite{Manton:1981mp,Ferrell:1987gf}. It would be interesting
to compare this analytic approximation method with a fully non-linear
NR simulation.

Motivation for the numerical modelling of charged
black holes also arises in the context of
high energy collisions. It is expected that trans-Planckian particle
collisions form black holes;
moreover, well above the fundamental Planck scale such processes should
be well described by general relativity
and other interactions should
become negligible~\cite{'tHooft:1987rb}, an idea poetically stated
as \textit{matter does not matter} for ultra high energy collisions~\cite{Choptuik:2009ww}. But is this expectation really correct?
Calculations of shock wave
collisions suggest that even though other interactions---say charge---may
become irrelevant in the ultra-relativistic limit, the properties of the
final black hole (and of the associated emission of gravitational
radiation) do depend on the amount of
charge carried by the colliding particles~\cite{Yoshino:2006dp,Gingrich:2006qh}. This issue can be clarified by the simulation of
high-energy collisions of charged black holes in the framework of NR 
and the subsequent comparison of the results to those obtained for
electrically neutral systems. Recent works in this direction include~\cite{Sperhake:2008ga,Shibata:2008rq,Sperhake:2009jz,Zilhao:2010sr,Witek:2010xi,Witek:2010az,Okawa:2011fv} 
for binary black holes and~\cite{Choptuik:2009ww} for boson stars. These, 
together with related incipient efforts to study gravity in higher-dimensional
space-times~\cite{Chesler:2008hg,Shibata:2009ad,Lehner:2010pn,Yoshino:2011zz,Cardoso:2012qm} illustrate 
recent applications of numerical simulations to shed light on problems beyond astrophysical settings.

In the context of astrophysics, charged black holes may be of
interest in realistic systems. First, a rotating black hole in an
external magnetic field will accrete charged particles up to
a given value, $Q=2B_0J$~\cite{Wald:1974np}. Thus it is conceivable
that astrophysical black holes
could have some (albeit rather small) amount of electrical charge. Then
it is of interest to understand the role of this charge
in the Blandford-Znajek mechanism~\cite{Blandford:1977ds},
which has been suggested for extracting spin energy from the hole,
or in a related mechanism capable of extracting energy from a moving black
hole~\cite{Palenzuela:2010nf,Palenzuela:2010xn} to power outflows from
accretion disk-fed black holes. NR simulations of
charged black holes interacting with matter and surrounding
plasma will enable us to study such effects.

Finally we note a variety of conceptual aspects that merit a more
detailed investigation of charged black-hole systems.
In head-on collisions with small velocity, the intuition borrowed from
Larmor's formula in Minkowski space suggests a steady growth of the emitted
power with the acceleration. However, it is by now well established
that for uncharged black holes the gravitational radiation strongly
peaks near the time of formation of a common apparent horizon.
Does the electromagnetic radiation emission follow a similar pattern?
And what is the relative fraction of electromagnetic to gravitational wave
emissions? Moreover, a non-head on collision of
charged non-spinning black holes will allow us to study, as the end
state, a (perturbed) Kerr-Newman geometry, which would be extremely
interesting: linearized perturbations around Kerr-Newman black holes do not decouple~\cite{MTB,Berti:2009kk}
and so far close to nothing is known about their properties. Among others, the stability of the Kerr-Newman metric
is an outstanding open issue.
Furthermore, it has been observed that the inspiral phase
of an orbiting black-hole-binary system can be well understood
via post-Newtonian methods~\cite{Blanchet2006} (see also 
e.g.~\cite{Boyle:2007ft, Sperhake:2011zz}).
The additional radiative channel opened by the presence of electric charge
provides additional scope to probe this observation.

With the above motivation in mind we here initiate the
numerical study of non-linear dynamics
of binary systems of RN black holes, building on
previous numerical evolutions of the Einstein-Maxwell system~\cite{Palenzuela:2008sf,Palenzuela:2009yr,Palenzuela:2009hx,Mosta:2009rr}.
For reasons of simplicity, we focus in this study on binary
systems for which initial data can be constructed by purely
analytic means~\cite{Brill:1963yv,Alcubierre:2009ij}:
head-on collisions, starting from rest, of non-spinning black holes with equal
charge-to-mass ratio. This implies in particular that the black holes
carry a charge of the same {\em sign} so that the electromagnetic force
will always be repulsive.
We will extract both gravitational and electromagnetic radiation
and monitor their behaviour as the charge-to-mass-ratio
parameter of the system is varied.

For this purpose, we present in
Sec.~\ref{sec:evol-eq} the evolution equations and the initial data
used. In Sec.~\ref{wave_extraction} the method for extraction of
gravitational and electromagnetic radiation is discussed. In Sec.~\ref{classical_expectations} we summarize our analytic calculations
and compare in Sec.~\ref{numerical_results}
their predictions with the numerical results. Throughout this work
greek ``spacetime indices'' run from $0$ to $3$ and latin ``spatial''
indices from $1$ to $3$.

\section{Evolution equations}
\label{sec:evol-eq}
In this paper we adopt the approach outlined
in~\cite{Komissarov:2007wk,Palenzuela:2009hx} to evolve the electro-vacuum
Einstein-Maxwell equations which incorporates suitably added additional
fields to ensure the evolution will preserve the constraints.
This amounts to considering an enlarged system of the form
\begin{equation}
  \label{eq:EFE}
  \begin{aligned}
    R_{\mu \nu} - \frac{R}{2} g_{\mu \nu} & = 8\pi T_{\mu \nu} \  ,\\
    \nabla_{\mu}\left( F^{\mu \nu} + g^{\mu\nu} \Psi
    \right) & = -\kappa n^{\nu} \Psi \ , \\
    \nabla_{\mu}\left(
    \star \!{}F^{\mu \nu} + g^{\mu\nu} \Phi
    \right) & = -\kappa n^{\nu} \Phi \ ,
  \end{aligned}
\end{equation}
where $\star \!{}F^{\mu \nu}$ denotes the Hodge dual of the
Maxwell-Faraday tensor $F^{\mu \nu}$, $\kappa$ is a constant and
$n^\mu$ the four-velocity of the Eulerian observer.  We recover the
standard Einstein-Maxwell system of equations when $\Psi = 0 = \Phi$.
With the scalar field $\Psi$ and pseudo-scalar $\Phi$ introduced in
this way, the natural evolution of this system drives $\Psi$ and $\Phi$
to zero (for positive $\kappa$), thus ensuring the magnetic and electric
constraints are controlled~\cite{Komissarov:2007wk,Palenzuela:2008sf}.
The electromagnetic stress-energy tensor takes the usual form
\begin{equation}
  \label{eq:Tmunu}
  T_{\mu \nu} = \frac{1}{4\pi} \left[ F_{\mu}{}^{\lambda} F_{\nu \lambda}
    - \frac{1}{4} g_{\mu \nu} F^{\lambda \sigma} F_{\lambda \sigma}
    \right] \ .
\end{equation}
%

\subsection{$3+1$ decomposition}
We employ a Cauchy approach so we introduce a $3+1$ decomposition of
all dynamical quantities. Concretely, we introduce the 3-metric
\begin{equation}
  \label{eq:3metric}
  \gamma_{\mu\nu} = g_{\mu \nu} + n_{\mu} n_{\nu} \ ,
\end{equation}
%
and decompose the Maxwell-Faraday tensor into the more familiar electric
and magnetic fields measured by the Eulerian observer moving with
four velocity $n^{\mu}$
%
\begin{equation}
  \begin{aligned}
    F_{\mu \nu} & = n_{\mu} E_{\nu} - n_{\nu} E_{\mu}
      + \epsilon_{\mu\nu\alpha\beta} B^{\alpha} n^{\beta}  \ ,\\
      \star \! F_{\mu \nu} & = n_{\mu} B_{\nu} - n_{\nu} B_{\mu}
      - \epsilon_{\mu\nu\alpha\beta} E^{\alpha} n^{\beta}  \ ,
  \end{aligned}
  \label{eq:faraday}
\end{equation}
where we use the convention $\epsilon_{1230} = \sqrt{-g}$,
$\epsilon_{\alpha \beta \gamma}
= \epsilon_{\alpha \beta \gamma \delta} n^{\delta}$,
$\epsilon_{123} = \sqrt{\gamma}$.

Writing the evolution equations in the BSSN form (see, e.g.,~\cite{Alcubierre:2008,Gourgoulhon:2007ue} for details), we have, for the
``gravitational'' part
\begin{equation}
  \begin{aligned}
    \tilde \gamma_{ij} & = \chi \gamma_{ij} \,, \quad \chi =\gamma^{-1/3} \ , \\
    \tilde{A}_{ij} & \equiv \chi\left(K_{ij}-\frac{\gamma_{ij}}{3}K\right)\ ,
  \end{aligned}
\end{equation}
\begin{equation}
  \label{eq:einstein-bssn}
  \begin{aligned}
    \left( \partial_t -  \mathcal{L}_\beta \right) \tilde \gamma_{ij} & =
        - 2 \alpha \tilde A_{ij}\  , \\
    \left( \partial_t -  \mathcal{L}_\beta \right) \chi  & =
        \frac{2}{3} \alpha \chi K\  , \\
    \left( \partial_t -  \mathcal{L}_\beta \right) K & =
        [\cdots] + 4 \pi \alpha (\rho + S)\  , \\
    \left( \partial_t -  \mathcal{L}_\beta \right) \tilde A_{ij} & =
        [\cdots] - 8 \pi \alpha \left(
          \chi S_{ij} - \frac{S}{3} \tilde \gamma_{ij}
        \right)\  , \\
    \left( \partial_t -  \mathcal{L}_\beta \right) \tilde \Gamma^i & =
        [\cdots] - 16 \pi \alpha \chi^{-1} j^i\  ,
    \quad \tilde{\Gamma}^i =\tilde{\gamma}^{jk}\tilde{\Gamma}^i_{jk} \ ,
  \end{aligned}
\end{equation}
where $ [\cdots] $ denotes the standard right-hand side of the BSSN equations in
the absence of source terms. For the case of the electromagnetic
energy-momentum tensor of Eqs.~(\ref{eq:Tmunu}), (\ref{eq:EFE}),
the source terms are given by
\begin{equation}
  \label{eq:source}
   \begin{aligned}
  \rho &\equiv T^{\mu \nu}n_{\mu}n_{\nu}
       = \frac{1}{8\pi} \left( E^2 + B^2 \right) \ , \\
  j_i  &\equiv -\gamma_{i\mu} T^{\mu \nu}n_{\nu}
       = \frac{1}{4\pi} \epsilon_{ijk} E^j B^k \ ,\\
  S_{ij} &\equiv \gamma^{\mu}{}_i \gamma^{\nu}{}_j T_{\mu \nu} \\
       &= \frac{1}{4\pi} \left[
         -E_i E_j - B_i B_j + \frac{1}{2} \gamma_{ij} \left( E^2 + B^2 \right)
         \right] \ ,
   \end{aligned}
\end{equation}
and $S\equiv \gamma^{ij}S_{ij}$.
The evolution of the electromagnetic fields is determined by
Eq.~(\ref{eq:EFE}) whose 3+1 decomposition
becomes~\cite{Mosta:2009rr}
\begin{widetext}
\begin{equation}
  \label{eq:maxwell-bssn}
  \begin{aligned}
    \left(
    \partial_t - \mathcal{L}_{\beta}
    \right) E^i & = \alpha K E^i + \epsilon^{ijk} \chi^{-1}
    \left[
    \tilde \gamma_{kl} B^l \partial_j \alpha 
    + \alpha \left(B^l \partial_j \tilde \gamma_{kl}
    + \tilde \gamma_{kl} \partial_j B^l
    - \chi^{-1} \tilde\gamma_{kl}B^l \partial_j \chi \right)
    \right]
    - \alpha \chi \tilde \gamma^{ij} \partial_j \Psi \ , \\
    \left(
    \partial_t - \mathcal{L}_{\beta}
    \right) B^i & = \alpha K B^i - \epsilon^{ijk} \chi^{-1}
    \left[
    \tilde \gamma_{kl} E^l \partial_j \alpha
    + \alpha \left(E^l \partial_j \tilde \gamma_{kl}
    + \tilde \gamma_{kl} \partial_j E^l
    - \chi^{-1} \tilde\gamma_{kl} E^l \partial_j \chi \right)
    \right]
    - \alpha \chi \tilde \gamma^{ij} \partial_j \Phi \ , \\
    \left(
    \partial_t - \mathcal{L}_{\beta}
    \right) \Psi & = -\alpha \nabla_i E^i - \alpha \kappa \Psi \ , 
    \qquad
    \left(
    \partial_t - \mathcal{L}_{\beta}
    \right) \Phi = -\alpha \nabla_i B^i - \alpha \kappa \Phi \ .
  \end{aligned}
\end{equation}
\end{widetext}
Here, $\mathcal{L}_{\beta}$ denotes the Lie derivative along
the shift vector $\beta^i$.
The Hamiltonian and momentum constraint are
\begin{equation}
  \label{eq:constraints}
  \begin{aligned}
    \mathcal{H} & \equiv {}^{3} \! R + K^2 - K^{ij}K_{ij} - 16 \pi \rho=0 \ , \\
    \mathcal{M}_i & \equiv D_j A_i{}^j - \frac{3}{2} A_i{}^j
        \chi^{-1} \partial_j \chi 
        - \frac{2}{3} \partial_i K -8\pi j_i = 0 \ ,
  \end{aligned}
\end{equation}
where $D_i$ is the covariant derivative associated with the three-metric
$\gamma_{ij}$.

\subsection{Initial data}
As already mentioned in the Introduction, we focus here 
on black-hole binaries with equal charge and mass colliding from
rest.
For these configurations, it is possible to construct initial data
using the Brill-Lindquist construction~\cite{Brill:1963yv} (see
also~\cite{Alcubierre:2009ij}). The main ingredients of this procedure are
as follows.

For a vanishing shift $\beta^i$, time symmetry implies
$K_{ij}=0$. Combined with the condition of an initially vanishing
magnetic field,
the magnetic constraint $D_i B^i=0$ and momentum constraint
are automatically satisfied. By further assuming the spatial metric
to be conformally flat
\begin{equation}
  \gamma_{ij} dx^i dx^j = \psi^4 \left( dx^2 + dy^2 + dz^2 \right) \ ,
  \label{eq:inigamma}
\end{equation}
the Hamiltonian constraint reduces to
\begin{equation}
  \triangle \psi = - \frac{1}{4} E^2 \psi^5 \ ,
\end{equation}
where $\triangle $ is the flat space Laplace operator. The electric
constraint, Gauss's law, has the usual form
\begin{equation}
  D_i E^{i} = 0 \ .
\end{equation}
%
Quite remarkably, for systems of black holes with equal charge-to-mass ratio,
these equations have known analytical solutions~\cite{Alcubierre:2009ij}.
For the special case of two black holes momentarily at rest with
``bare masses'' $m_1$, $m_2$ and ``bare charges'' $q_1$, $q_2=q_1 m_2/m_1$
this analytic solution is given by
\begin{equation}
  \begin{aligned}
    \psi^2 & = \left( 1 + \frac{m_1}{2|\vec x - \vec x_1|}
      + \frac{m_2}{2|\vec x - \vec x_2|} \right)^2 \\
    & \quad  - \frac{1}{4} \left( \frac{q_1}{|\vec x - \vec x_1|}
      + \frac{q_2}{|\vec x - \vec x_2|} \right)^2  \ , \\
    E^i & = \psi^{-6} \left( q_1
        \frac{ (\vec x - \vec x_1)^i }{|\vec x - \vec x_1|^{3}}
      + q_2 \frac{ (\vec x - \vec x_2)^i }{|\vec x - \vec x_2|^{3}}
    \right) \ ,
  \end{aligned}
  \label{eq:ini_psi_E}
\end{equation}
where $\vec x_i$ is the coordinate location of the $i$th ``puncture''.\footnote{We note that this foliation, in isotropic coordinates, only covers the outside of the external horizon.}

The initial data are thus completely specified in terms of the independent
mass and charge parameters $m_1$, $m_2$, $q_1$ and the initial
coordinate separation $d$ of the holes. These uniquely determine the
remaining charge parameter $q_2$ via the condition
of equal charge-to-mass ratio. In this study we always choose $m_1=m_2$
and, without loss of generality, position the two holes symmetrically
around the origin such that $z_1=d/2=-z_2$. The resulting initial
three metric $\gamma_{ij}$
follows from Eqs.~(\ref{eq:inigamma}), (\ref{eq:ini_psi_E})
while the extrinsic curvature $K_{ij}$ and magnetic field $B^i$
vanish on the initial slice.

Finally, the time evolution of the fields is determined by Eqs.~(\ref{eq:einstein-bssn}) and (\ref{eq:maxwell-bssn}). We use the same gauge conditions and outer boundary conditions for the BSSN variables as used in vacuum simulations~\cite{Alcubierre:2002kk}. As outer boundary condition for the electric and magnetic fields we have imposed a falloff as $1/r^2$---from~\eqref{eq:ini_psi_E}. For the additional scalar fields a satisfactory behaviour is observed by imposing a falloff as $1/r^3$
(which is the expected falloff rate from dimensional grounds).

\section{Wave Extraction}
\label{wave_extraction}

For a given set of initial parameters $m_1=m_2$, $q_1=q_2$, $d$,
the time evolution provides us
with the spatial metric $\gamma_{ij}$, the extrinsic curvature $K_{ij}$
as well as the electric and magnetic fields $E^i$, $B^i$ as functions
of time. These fields enable us to extract the gravitational
and electromagnetic radiation as follows.

For the gravitational wave signal we calculate the Newman-Penrose
scalar $\Psi_4$ defined as
\begin{equation}
  \label{eq:Psi4}
  \Psi_4 \equiv C_{\alpha\beta\gamma\delta}
      k^{\alpha} \bar m^{\beta} k^\gamma \bar m^{\delta} \ ,
\end{equation}
where $C_{\alpha\beta\gamma\delta}$ is the Weyl tensor and $k$, $\bar m$
are part of a null tetrad $l,k,m,\bar m$ satisfying $-l \cdot k = 1 =
m \cdot \bar m$; all other inner products vanish. In practice
$l$, $k$ and $m$ are constructed from an orthonormal triad $u, v, w$
orthogonal to the unit timelike vector $n^{\mu}$:
\begin{equation}
\begin{aligned}
l^{\alpha} & = \frac{1}{\sqrt{2}} \left( n^{\alpha} + u^{\alpha} \right) \ , \\
k^{\alpha} & = \frac{1}{\sqrt{2}} \left( n^{\alpha} - u^{\alpha} \right) \ , \\
m^{\alpha} & = \frac{1}{\sqrt{2}} \left( v^{\alpha} + i w^{\alpha} \right) \ .
\end{aligned}
\end{equation}
We refer the interested reader to~\cite{Sperhake:2006cy} for more
details about the numerical implementation and~\cite{Lehner:2007ip}
for a review of the formalism; here we merely note that asymptotically
the triad vectors $u,~v,~w$ behave as the unit radial, polar and
azimuthal vectors $\hat r,~\hat{\theta},~\hat{\phi}$.

Similarly, we extract the electromagnetic wave signal in the form
of the scalar functions, $\Phi_1$ and $\Phi_2$~\cite{Newman:1961qr},
defined as
\begin{align}
  \label{eq:Phi1}
  \Phi_1 & \equiv \frac{1}{2} F_{\mu \nu} \left(
    l^{\mu} k^{\nu} + \bar m^{\mu} m^{\nu}
  \right) \ , \\
  \label{eq:Phi2}
  \Phi_{2} & \equiv F_{\mu \nu} \bar m^{\mu} k^{\nu} \ .
\end{align}
For outgoing waves at infinity, these quantities behave as
\begin{equation}
  \label{eq:Phi_asympt}
  \Phi_1 \sim \frac{1}{2}\left(
    E_{\hat r} + i B_{\hat r}
    \right)\ , \quad \Phi_2 \sim E_{\hat \theta} - i E_{\hat \phi} \ .
\end{equation}

At a given extraction radius $R_\mathrm{ex}$, we perform a multipolar
decomposition by projecting $\Psi_4$, $\Phi_1$ and $\Phi_2$ onto spherical
harmonics of spin weight $s=-2$, $0$ and $-1$ respectively:
\begin{align}
  \Psi_4(t, \theta, \phi) & =
      \sum_{l,m} \psi^{lm}(t) Y_{lm}^{-2}(\theta,\phi) \ , \\
  \Phi_1(t, \theta, \phi) & =
      \sum_{l,m} \phi_{1}^{lm}(t) Y_{lm}^{0}(\theta,\phi) \ ,
      \label{eq:multipole_Phi1} \\
  \Phi_2(t, \theta, \phi) & =
      \sum_{l,m} \phi_{2}^{lm}(t) Y_{lm}^{-1}(\theta,\phi) \ .
\end{align}
In terms of these multipoles, the radiated flux and energy is given by the
expressions~\cite{Newman:1961qr} 
\begin{align}
  \label{eq:GW-flux}
  F_{\rm GW} & = \frac{d E_{\rm GW}}{dt} =
      \lim_{r\to\infty} \frac{r^2}{16 \pi} \sum_{l,m}
      \left| \int_{-\infty}^t dt' \psi^{lm} (t') \right|^2 \ , \\
  F_{\rm EM} & = \frac{d E_{\rm EM}}{dt} =
      \lim_{r\to\infty} \frac{r^2}{4 \pi} \sum_{l,m} 
      \left|  \phi^{lm}_{2} (t) \right|^2 \ . \label{eq:EM-flux}
\end{align}
As is well known from simulations of uncharged black-hole binaries,
initial data obtained from the Brill-Lindquist construction contain
``spurious'' radiation, which is an artifact of the conformal-flatness
assumption. In calculating properties of the radiation, we account for
this effect by starting the integration of the radiated flux
in Eqs.~(\ref{eq:GW-flux}), (\ref{eq:EM-flux}) at some finite time $\Delta t$
after the start of the simulation, thus allowing the spurious pulse
to first radiate off the computational domain. In practice, we obtain
satisfactory results by choosing $\Delta t = R_{\rm ex}+50~M$.
Because the physical radiation is very weak for both the gravitational
and electromagnetic channel in this early infall stage, the error
incurred by this truncation is negligible compared with the uncertainties
due to discretization; cf.~Sec.~\ref{sec:fluxes}.

\section{Analytic predictions}
\label{classical_expectations}
Before discussing in detail the results of our numerical simulations,
it is instructive to discuss the behaviour of the binary system as
expected from an analytic approximation. Such an analysis not only serves
an intuitive understanding of the binary's dynamics, but also provides
predictions to compare with the numerical results presented below.

For this purpose we consider the electrodynamics of a system of
two equal point charges in a Minkowski background spacetime. As in
the black-hole case, we denote by $q_1=q_2\equiv Q/2$ and $m_1=m_2\equiv M/2$
the electric charge and mass of the particles which are initially
at rest at position $z=\pm d/2$.

It turns out to be useful to first consider point charges in Minkowski spacetime
in the static limit. The expected behaviour
of the radial component of the resulting electric field is given
by~\cite{Jackson1998Classical}
\begin{equation}
  \label{eq:Er_asympt}
  E_{\hat r} = 4\pi\sum_{l,m} \frac{l+1}{2l+1} q_{lm}
      \frac{Y_{lm}(\theta,\varphi)}{r^{l+2}} \ ,
\end{equation}
which for a system of two charges of equal magnitude at $z=\pm d/2$ becomes
\begin{equation}
  \label{eq:Er_2q}
  E_{\hat r} \simeq \sqrt{4\pi} Q \frac{Y_{00}}{r^2} 
     + \sqrt{\frac{9\pi}{20}} Q d^2 \frac{Y_{20}}{r^4} \ .
\end{equation}
The dipole vanishes in this case due to the reflection symmetry across
$z=0$. This symmetry is naturally preserved during the time evolution
of the two-charge system. Furthermore, the total electric charge $Q$
is conserved so that the leading-order behaviour of the
electromagnetic radiation is given by variation of the electric
quadrupole, just as for the gravitational radiation.
Notice that in principle other radiative contributions can arise
from the accelerated motion of the charged black holes.
From experience with gravitational radiation generated in the collision
of electrically neutral black-hole binaries, however, we expect
this ``Bremsstrahlung'' to be small in comparison with the merger
signal and hence ignore its contributions in this simple approximation.
The good agreement with the numerical results presented in the
next section bears out the validity of this {\em quadrupole approximation}.
In consequence, it appears legitimate to regard the ``strength'' of the
collision and the excitation of the black-hole ringdown to be
purely kinematic effects.

An estimate for the monopole and quadrupole amplitudes in the limit
of two static point charges is then obtained from inserting the
radial component of the electric field (\ref{eq:Er_2q}) into
the expression (\ref{eq:Phi_asympt}) for $\Phi_1$ and its
multipolar decomposition (\ref{eq:multipole_Phi1})
\begin{eqnarray}
  r^2\phi_1^{00}&=&\sqrt{\pi} Q\approx 1.77 Q\,,\label{eq:mono}\\
  r^4\phi_1^{20}&=&\sqrt{\frac{9\pi}{80}} Q d^2\approx 0.59 Qd^2\,.
     \label{eq:dipole}
\end{eqnarray}
The expectation is that these expressions provide a good approximation for
the wave signal during the early infall stage when the black holes are
moving with small velocities. Equation~(\ref{eq:mono}) should also provide a
good approximation for $\phi_1^{00}$ after the merger and ringdown
whereas the quadrupole $\phi_1^{20}$ should eventually approach zero
as a single merged hole corresponds to the case $d=0$ in
Eq.~(\ref{eq:dipole}).

To obtain analytic estimates for the collision time and the
emitted radiation, we need to describe the dynamic behaviour of the
two point charges. Our starting point for this discussion is the
combined gravitational and electromagnetic potential energy for two charges
$i=1,\,2$ in Minkowski spacetime
with mass and charge $m_i$, $q_i$ at distance $r$ from each other
\begin{equation}
  V=-\frac{Gm_1m_2}{r}+\frac{1}{4\pi\epsilon_0}\frac{q_1q_2}{r}\ .
\end{equation}
For the case of two charges
with equal mass and charge $m_i=M/2$,
$q_i=Q/2$ and
starting from rest at $z_0=\pm d/2$,
conservation of energy implies
\begin{equation}
  M\dot{z}^2-\frac{M^2{\cal B}}{4z}=-\frac{M^2{\cal B}}{2d} \ , \label{eqm2}
\end{equation}
where we have used units with $G=4\pi \epsilon_0 = 1$ and
\begin{equation}
  {\cal B}\equiv 1-Q^2/M^2\ .
\end{equation}
The resulting equation of motion for $z(t)$ is obtained
by differentiating Eq.~(\ref{eqm2}) which results in
\begin{equation}
  M\ddot{z}=-\frac{M^2}{8z^2}+\frac{Q^2}{8z^2}=
      -M^2\frac{{\cal B}}{8z^2}\ . \label{eqm1}
\end{equation}
An estimate for the time for collision follows from
integrating Eq.~(\ref{eqm2}) over $z\in [d/2,0]$
\begin{equation}
  \left(\frac{t_{\rm collision}}{M}\right)^2
      =\frac{\pi^2 d^3}{2^3M^3 {\cal B}}\ . \label{eq:timecollision}
\end{equation}

From the dynamic evolution of the system we can derive an approximate
prediction for the electromagnetic radiation by evaluating the (traceless)
electric quadrupole tensor $Q_{ij}=\int d^3\vec{x} \rho(\vec{x})(3x_ix_j-r^2\delta_{ij})$~\cite{Jackson1998Classical}.
%
%
In terms of
this quadrupole tensor, the total power radiated is given by~\cite{Jackson1998Classical}
\begin{equation}
F_{\rm EM}=\sum_{ij}\frac{1}{4\pi\epsilon_0}
      \frac{1}{360c^5}\dddot{Q}_{ij}^2\ .\label{eq:power}
\end{equation}
%
%
For clarity we have reinstated the factors $4\pi \epsilon_0$ and $c^5$ here.
Using
\begin{equation}
\frac{d^3}{dt^3}(z^2)=6\dot{z}\ddot{z}+2z\dddot{z}\ ,
\end{equation}
and the equations of motion (\ref{eqm2}), (\ref{eqm1}) we find
\begin{equation}
F_{\rm EM}=\frac{{\cal B}^3M^3Q^2(1/z-2/d)}{1920z^4}\ .
\end{equation}
Using $\int dt (\ldots ) = \int dz/\dot{z} (\ldots)$, we can evaluate
the time integral up to some cutoff separation, say
$z_{\rm min} = \alpha_b b$, where $b$ is the horizon radius of the
initial black hole,
$b=M(1+\sqrt{{\cal B}})/2$ and $\alpha_b={\cal O}(1)$ is a constant. This
gives,
\begin{widetext}
  \begin{equation}
    \label{EMquadprediction}
    \frac{E^{\rm EM}_{\rm rad}}{M} =
    {\cal B}^{5/2}M^{3/2}Q^2 
     \frac{(d-2\alpha_b b)^{3/2}
      (15d^2+24d\alpha_b b+32\alpha_b^2b^2)}{50400(d\alpha_b b)^{7/2}}\ .
  \end{equation}

  Emission of gravitational radiation follows from the quadrupole formula, which
  is a numerical factor $4$ times larger, and where the charge is be replaced by
  the mass,
  \begin{equation}
    \label{GWquadprediction}
    \frac{E^{\rm GW}_{\rm rad}}{M} = {\cal B}^{5/2}M^{7/2}
     \frac{(d-2\alpha_b b)^{3/2}
      (15d^2+24d\alpha_b b+32\alpha_b^2b^2)}{12600(d\alpha_b b)^{7/2}}\ .
  \end{equation}
\end{widetext}
For $Q=0, \alpha_b=1, d=\infty$ we thus obtain
\begin{equation}
  \frac{E^{\rm GW}_{\rm rad}}{M}=\frac{1}{840}\sim 0.0012 \ ,
\end{equation}
in agreement to within a factor of $2$ with numerical simulations (see~\cite{Witek:2010xi} and Table~\ref{tab:runs} below; the agreement could
be improved by assuming $\alpha_b\sim 1.3$). As a general result
of this analysis we find in this approximation,
\begin{equation}
  \frac{E^{\rm EM}_{\rm rad}}{E^{\rm GW}_{\rm rad}}=\frac{Q^2}{4M^2}\ .
\label{prediction_ratio}
\end{equation}
For non-extremal holes $Q<M$, our analytic considerations
therefore predict that the energy emitted in
electromagnetic radiation is at most $25\%$ of the energy lost in
gravitational radiation. As we shall see below, this turns out to be
a remarkably good prediction for the results obtained from
fully numerical simulations.

\section{Numerical Results}
\label{numerical_results}
%
\begin{table*}[ht]
  \centering
  \caption{Grid structure in the notation of Sec.~II E of~\cite{Sperhake:2006cy}, coordinate distance
  $d/M$, proper horizon-to-horizon
  distance $L/M$, charge $Q/M$, gravitational ($E_{\rm
  rad}^{\mathrm{GW}}$) and electromagnetic ($E_{\rm rad}^{\mathrm{EM}}$)
  radiated energy for our set of simulations. The radiated energy has
  been computed using
  only the $l=2$, $m=0$ mode; the energy contained higher-order
  multipoles such as $l=4$, $m=0$ is negligible for all configurations.
  \label{tab:runs}}
  \begin{tabular*}{\textwidth}{@{\extracolsep{\fill}}lcccccc}
    \hline
    \hline
    Run     &      Grid                                       &  $d/M$  & $L/M$ & $Q/M$ & $E_{\rm rad}^{\mathrm{GW}}$    & $E_{\rm rad}^{\mathrm{EM}}$ \\
    \hline
    d08q00   & $\{(256,128,64,32,16,8)\times(2,1,0.5), 1/80\}$ & 8.002   & 11.56 & 0     & $5.1\times10^{-4}$ &  --            \\
    d08q03   & $\{(256,128,64,32,16,8)\times(2,1,0.5), 1/80\}$ & 8.002   & 11.60 & 0.3   & $4.5\times10^{-4}$ & $1.3\times10^{-5}$ \\
    d08q04   & $\{(256,128,64,32,16,8)\times(2,1,0.5), 1/80\}$ & 8.002   & 11.65 & 0.4   & $4.0\times10^{-4}$ & $2.1\times10^{-5}$ \\
    d08q05c  & $\{(256,128,64,32,16,8)\times(2,1,0.5), 1/64\}$ & 8.002   & 11.67 & 0.5   & $3.3\times10^{-4}$ & $2.7\times10^{-5}$ \\
    d08q05m  & $\{(256,128,64,32,16,8)\times(2,1,0.5), 1/80\}$ & 8.002   & 11.70 & 0.5   & $3.4\times10^{-4}$ & $2.7\times10^{-5}$ \\
    d08q05f  & $\{(256,128,64,32,16,8)\times(2,1,0.5), 1/96\}$ & 8.002   & 11.67 & 0.5   & $3.4\times10^{-4}$ & $2.7\times10^{-5}$ \\
    d08q055  & $\{(256,128,64,32,16,8)\times(2,1,0.5), 1/80\}$ & 8.002   & 11.70 & 0.55  & $3.0\times10^{-4}$ & $2.89\times10^{-5}$ \\
    d08q06   & $\{(256,128,64,32,16,8)\times(2,1,0.5), 1/80\}$ & 8.002   & 11.75 & 0.6   & $2.6\times10^{-4}$ & $2.97\times10^{-5}$ \\
    d08q07   & $\{(256,128,64,32,16,8)\times(2,1,0.5), 1/80\}$ & 8.002   & 11.87 & 0.7   & $1.8\times10^{-4}$ & $2.7\times10^{-5}$ \\
    d08q08   & $\{(256,128,64,32,16,8)\times(2,1,0.5), 1/80\}$ & 8.002   & 12.0  & 0.8   & $9.8\times10^{-5}$ & $1.8\times10^{-5}$ \\
    d08q09   & $\{(256,128,64,32,16,8)\times(2,1,0.5), 1/80\}$ & 8.002   & 12.3  & 0.9   & $2.6\times10^{-5}$ & $5.5\times10^{-6}$ \\
    d08q098cc & $\{(256,128,64,32,16,8)\times(2,1,0.5), 1/64\}$ & 8.002   & 12.3  & 0.98  & $7.0\times10^{-7}$ & $2.1\times10^{-7}$ \\
    d08q098c & $\{(256,128,64,32,16,8)\times(2,1,0.5), 1/80\}$ & 8.002   & 13.1  & 0.98  & $4.3\times10^{-7}$ & $1.4\times10^{-7}$ \\
    d08q098m & $\{(256,128,64,32,16,8)\times(2,1,0.5), 1/96\}$ & 8.002   & 13.1  & 0.98  & $3.4\times10^{-7}$ & $1.0\times10^{-7}$ \\
    d08q098f & $\{(256,128,64,32,16,8)\times(2,1,0.5), 1/112\}$ & 8.002 & 13.0  & 0.98  & $4.0\times10^{-7}$ & $9.5\times10^{-8}$ \\
    d08q098ff & $\{(256,128,64,32,16,8)\times(2,1,0.5), 1/128\}$ & 8.002 & 13.0 & 0.98  & $4.05\times10^{-7}$ & $8.75\times10^{-8}$ \\
    d08q098fff & $\{(256,128,64,32,16,8)\times(2,1,0.5), 1/136\}$ & 8.002 & 13.1 & 0.98  & $3.73\times10^{-7}$ & $8.41\times10^{-8}$ \\
    d16q00   & $\{(256,128,64,32,16)\times(4,2,1,0.5), 1/64\}$ & 16.002  & 20.2  & 0     & $5.5\times10^{-4}$ & --                \\
    d16q05   & $\{(256,128,64,32,16)\times(4,2,1,0.5), 1/64\}$ & 16.002  & 20.3  & 0.5   & $3.6\times10^{-4}$ & $2.9\times10^{-5}$ \\
    d16q08   & $\{(256,128,64,32,16)\times(4,2,1,0.5), 1/80\}$ & 16.002  & 20.7  & 0.8   & $1.05\times10^{-4}$ & $1.9\times10^{-5}$ \\
    d16q09   & $\{(256,128,64,32,16)\times(4,2,1,0.5), 1/80\}$ & 16.002  & 21.0  & 0.9   & $2.7\times10^{-5}$  & $5.9\times10^{-6}$ \\
 
    \hline
    \hline
  \end{tabular*}
\end{table*}
\begin{figure*}[ht]
\centering
\includegraphics[width=0.45\textwidth]{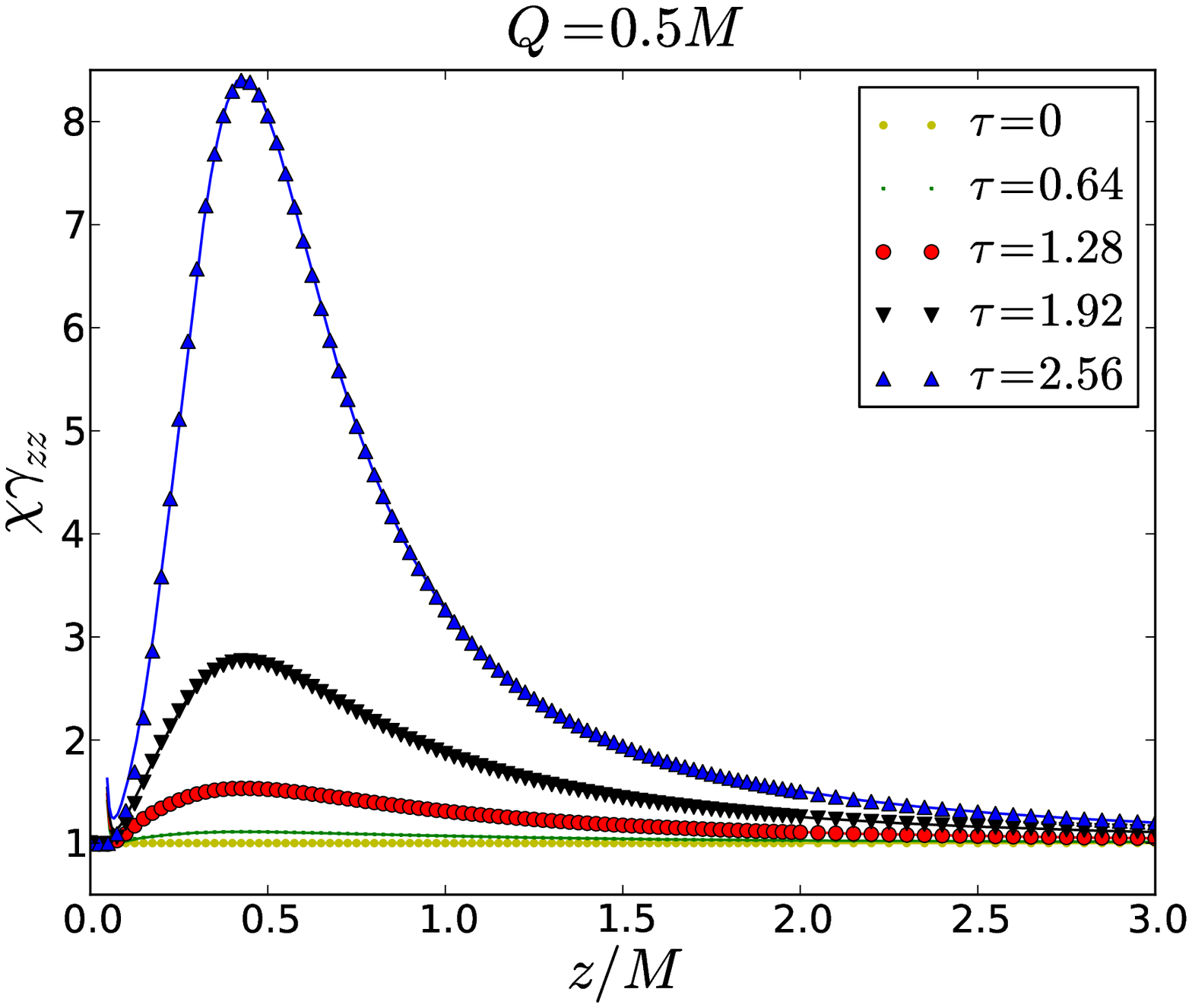}
\includegraphics[width=0.45\textwidth]{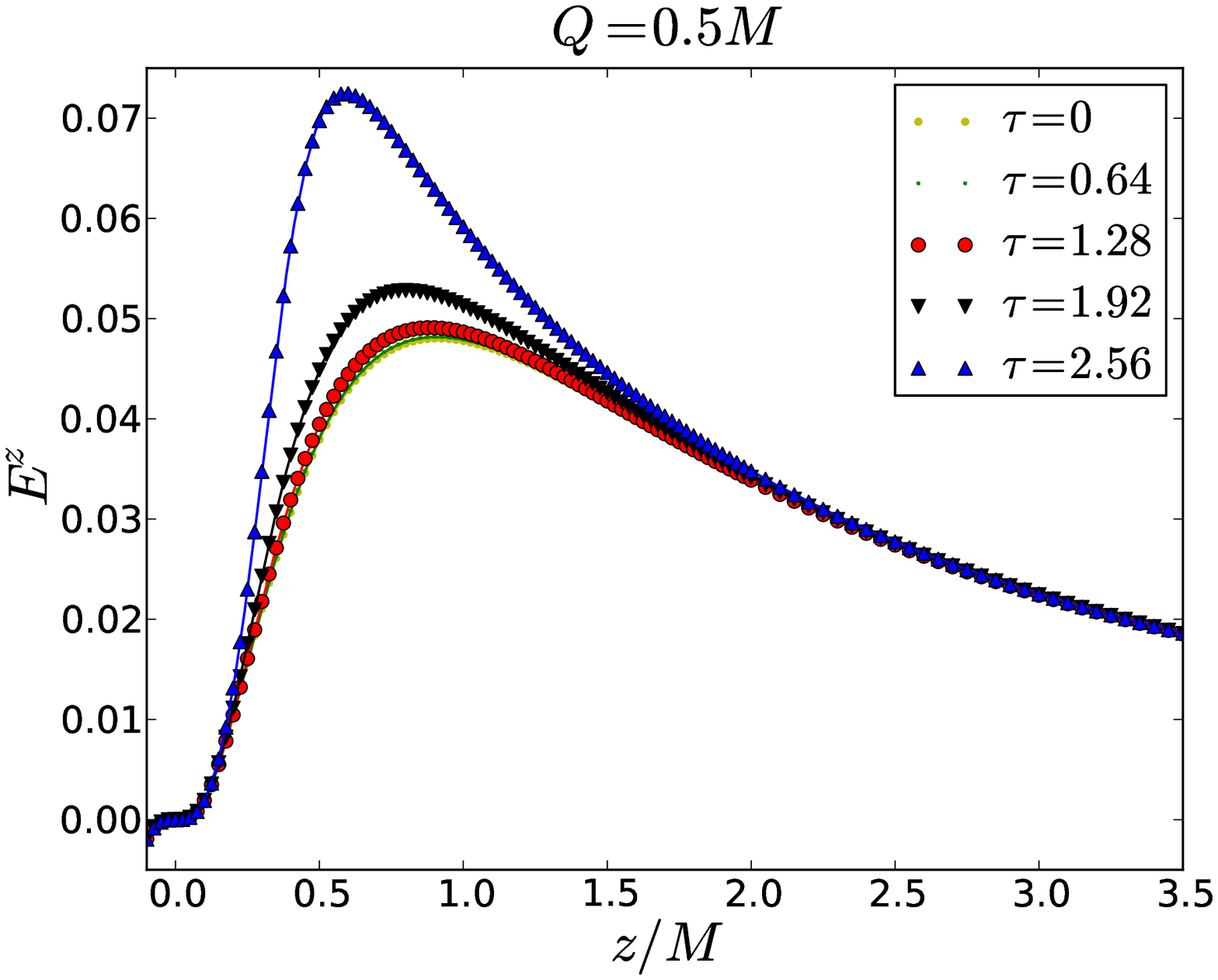}
\caption[]{The numerical profiles for $\gamma_{zz}$ and $E^z$ (symbols)
  obtained in geodesic slicing at various times $\tau$ are compared
  with the semi-analytic results (lines).}
\label{fig:geo_slice} 
\end{figure*}
The numerical integration of the Einstein-Maxwell equations
(\ref{eq:einstein-bssn}), (\ref{eq:maxwell-bssn})
has been performed using fourth-order spatial discretization
with the \textsc{Lean} code, originally presented
in~\cite{Sperhake:2006cy} for vacuum spacetimes.  \textsc{Lean} is
based on the \textsc{Cactus} Computational toolkit~\cite{cactus}, the
\textsc{Carpet} mesh refinement package~\cite{Schnetter:2003rb,carpet}
and uses \textsc{AHFinderDirect} for tracking apparent
horizons~\cite{Thornburg:2003sf,Thornburg:1995cp}. For further
details of the numerical methods see Ref.~\cite{Sperhake:2006cy}.

The initial parameters as well as the grid setup and the radiated
gravitational and electromagnetic wave energy for our
set of binary configurations is listed
in Table~\ref{tab:runs}. All binaries start from rest with a coordinate
distance $d/M\simeq 8$ or $d/M\simeq 16$ while the charge-to-mass
ratio has been varied from $Q/M=0$ to $Q/M=0.98$.
Note that identical coordinate separations of the punctures
for different values of the charge $Q/M$
correspond to different horizon-to-horizon proper distances.
This difference is expected and in fact analysis of the RN solution
predicts a divergence of the proper distance in the limit
$Q/M\rightarrow 1$.

\subsection{Code tests}
\label{sec:tests}
Before discussing the obtained results in more
detail, we present two tests to validate the performance of our
numerical implementation of the evolution equations. (i) Single
black-hole evolutions in {\em geodesic slicing} which is known to
result in numerical instabilities after relatively short times but
facilitates direct comparison with a semi-analytic solution and
(ii) Convergence analysis of the radiated quadrupole waveforms
for simulation d08q05 of Table~\ref{tab:runs}.

The geodesic slicing condition is enforced by setting the gauge functions
to $\alpha=1$, $\beta^i=0$ throughout the evolution. The space part
of the Reissner-Nordstr{\"o}m solution in isotropic coordinates
is given by Eq.~(\ref{eq:inigamma}) with a conformal factor~\cite{Graves:1960zz,Reimann:2003zd}
\begin{equation}
  \psi^2 = \left( 1+\frac{M}{2r} \right)^2 - \frac{Q^{2}}{4r^2}\ .
  \label{eq:conformalfactor}
\end{equation}
The time evolution of this solution is not known in closed analytic
form, but the resulting metric components can be constructed
straightforwardly via a simple integration procedure, cf.~Appendix~\ref{sec:geo}.
As expected, we find a time evolution in this gauge to become
numerically unstable at times $\tau$ of a few $M$.
Before the breaking
down of the evolution, however, we can safely compare the numerical and
``analytical'' solutions. This comparison is shown in Fig.~\ref{fig:geo_slice}
for the $\gamma_{zz}$ component of the spatial metric and the $E^z$
component of the electric field
and demonstrates excellent agreement between the semi-analytic
\begin{figure*}[htbp]
\centering
\includegraphics[width=0.45\textwidth]{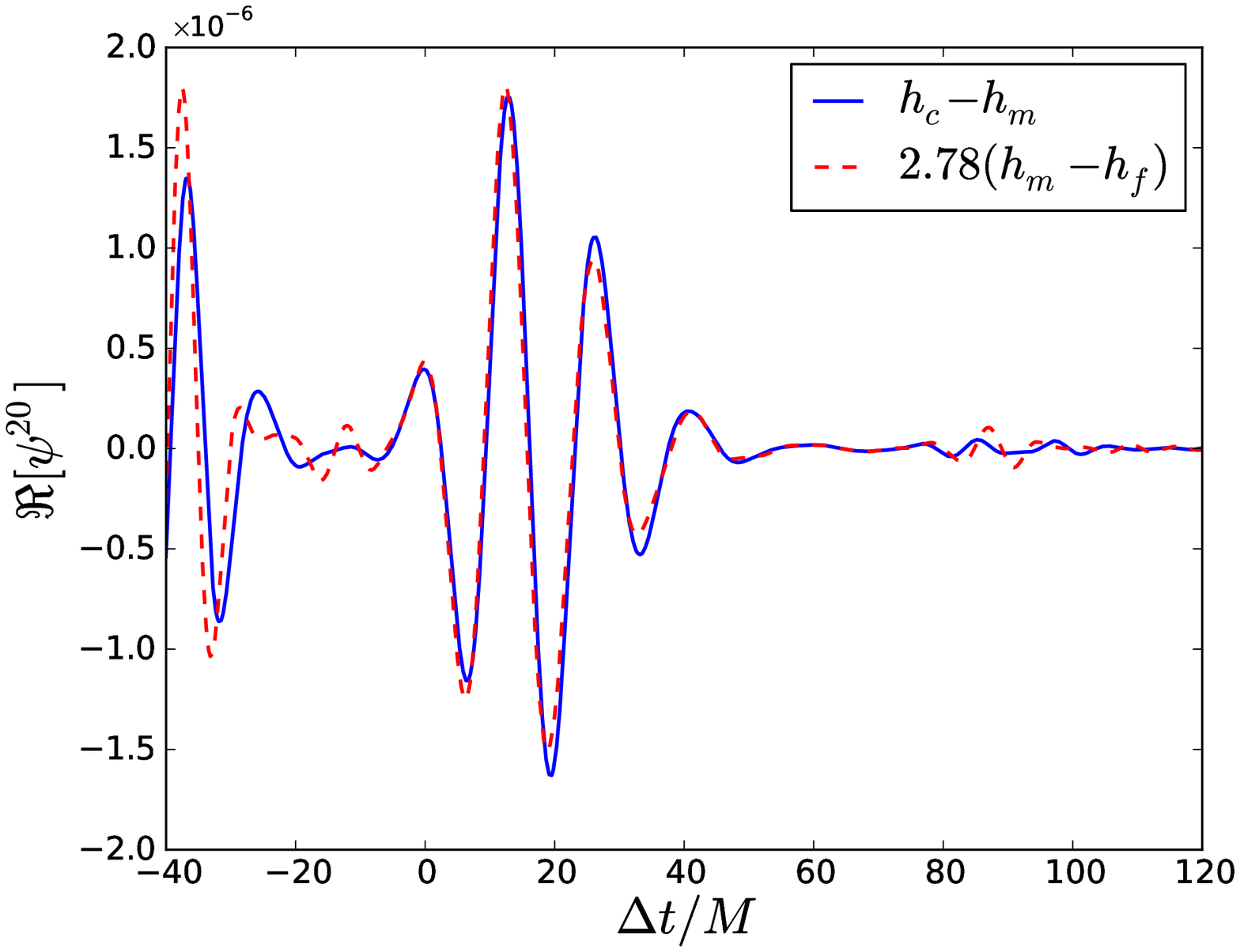}
\includegraphics[width=0.45\textwidth]{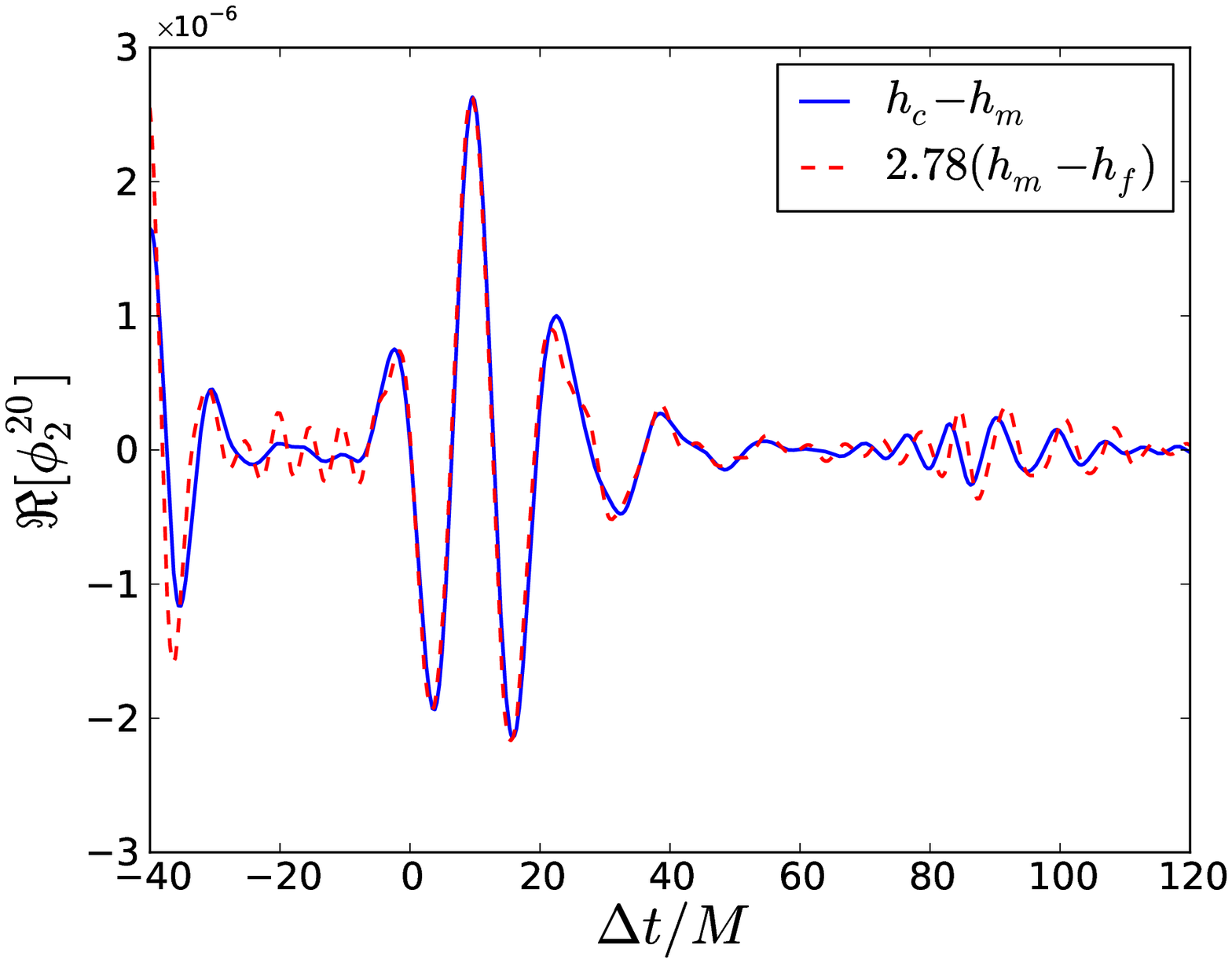}
\caption[]{Convergence analysis for simulation d08q05 of Table~\ref{tab:runs}
  with resolutions $h_c= M/64$, $h_m=M/80$ and $h_f=M/96$.
  The panels show differences of the $(2,0)$ multipoles of
  the real parts of $\Psi_4$ (left) and $\Phi_2$ (right)
  extracted at $R_{\rm ex}=100~M$;
  in each case, the high-resolution differences have been rescaled
  by a factor 2.78 as expected for fourth-order convergence.}
\label{fig:convergence}
\end{figure*}
and numerical results.

For the second test, we have evolved model d08q05
using three different resolutions as listed in Table~\ref{tab:runs}
and extracted the gravitational and electromagnetic quadrupole $(l=2,m=0)$
at $R_{\rm ex}=100~M$. For fourth-order convergence, we expect the
differences between the higher resolution simulations to be a factor
$2.78$ smaller than their coarser resolution counterparts. The numerically
obtained differences are displayed with the corresponding
rescaling in Fig.~\ref{fig:convergence}. Throughout the physically
relevant part of the waveform, we observe the expected fourth-order
convergence. Only the spurious initial radiation (cf.~the discussion
at the end of Sec.~\ref{wave_extraction}) at early times
$\Delta t \lesssim -20$ in the figure exhibits convergence closer to
second order, presumably a consequence of high-frequency noise contained
in this spurious part of the signal. From Richardson extrapolation
of our results we estimate the truncation error of the radiated waves
to be about $1~\%$.
The error due to extraction at finite radius, on the other hand, is estimated to be 2 \% at $R_{\rm ex}=100~M$.

\subsection{Collisions of two black holes: the ``static'' components and infall time}

We start the discussion of our results with the behaviour of the
gravitational and electromagnetic multipoles when the system is
\begin{figure*}[htbp]
\centering
\includegraphics[width=0.45\textwidth]{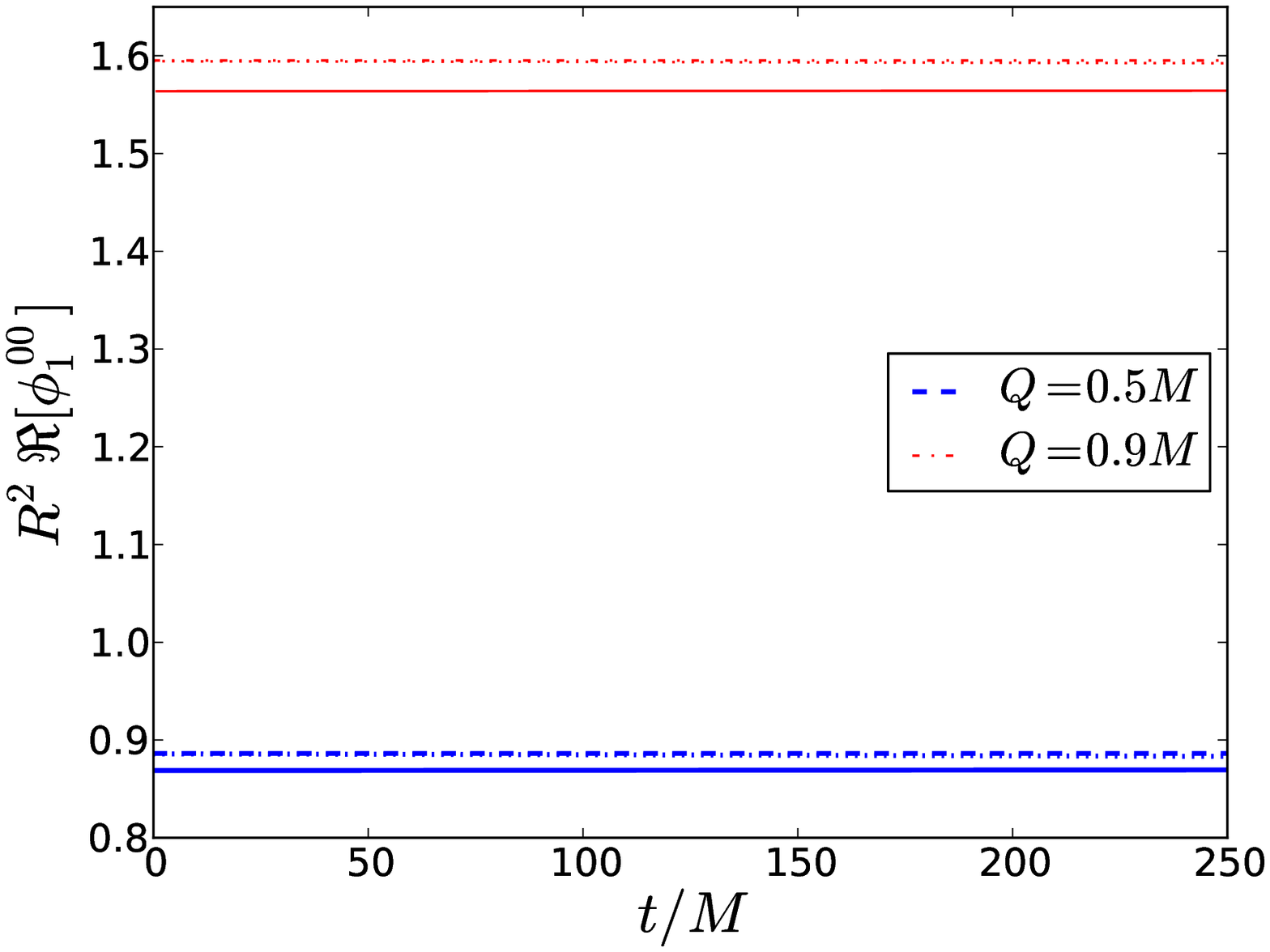}
\includegraphics[width=0.45\textwidth]{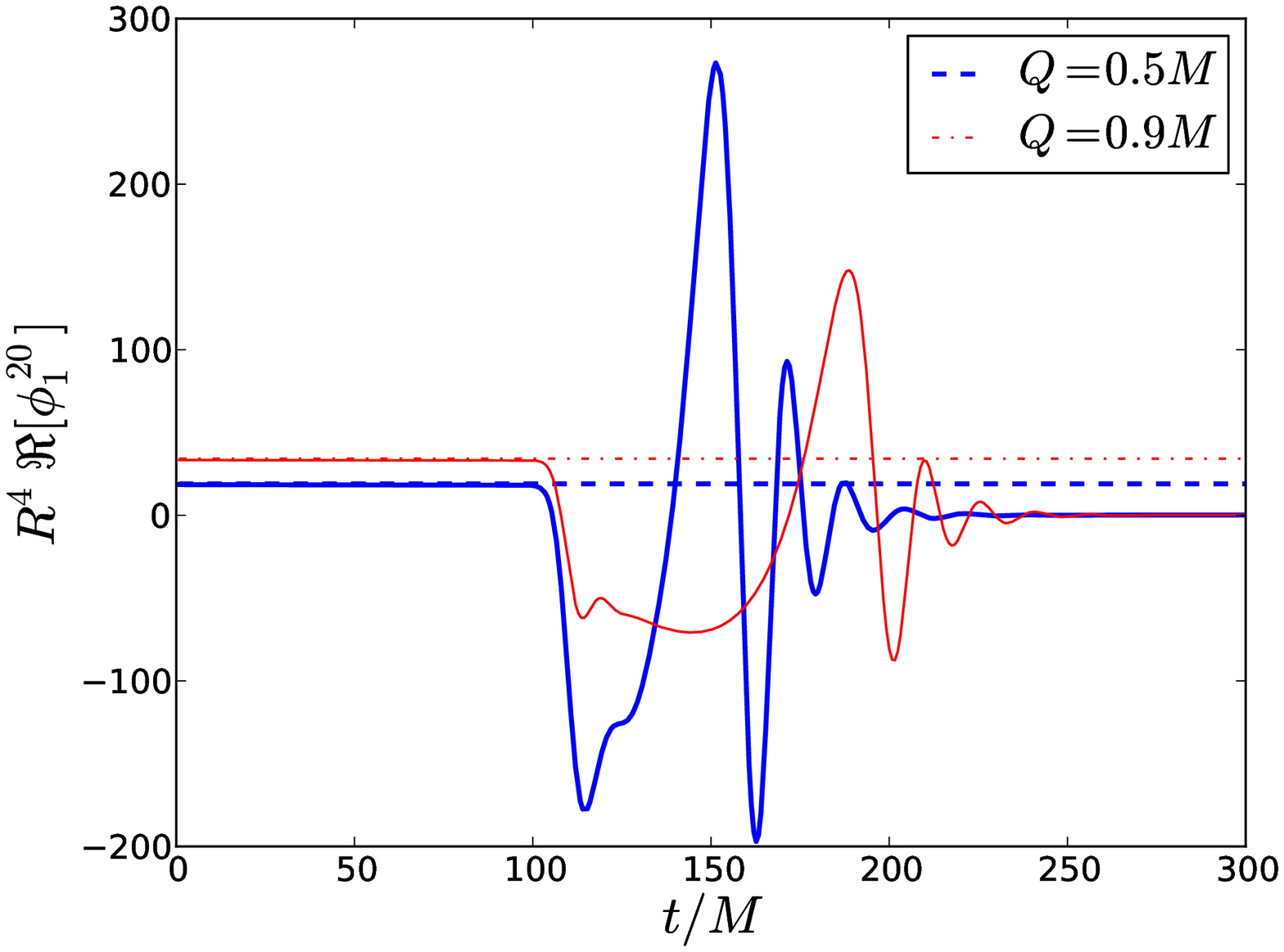}
\caption[]{Monopole $\phi_1^{00}$ (left) and quadrupole $\phi_1^{20}$ (right) 
  of the radial part of the electromagnetic field $\Phi_1$
  extracted at $R_{\rm ex}=100~M$ for simulation d08q05 of
  Table~\ref{tab:runs}. The dashed curves
  show the predictions of Eqs.~(\ref{eq:mono}), (\ref{eq:dipole})
  at $R=\infty$ in the static limit. For the monopole case, we also added the curves obtained by extrapolating the results to infinite extraction radius; these curves---dotted lines---essentially overlap with the predictions from Eq.~(\ref{eq:mono}). 
  \label{fig:multipoles}}
\end{figure*}
in a nearly static configuration, i.e.~shortly after the start of
the simulation and at late stages after the ringdown of the post-merger
hole. At these times, we expect our analytic predictions
(\ref{eq:mono}), (\ref{eq:dipole}) for the monopole and dipole of the
electromagnetic field to provide a rather accurate description.
Furthermore, the total spacetime charge $Q$ is conserved throughout
the evolution, so that the monopole component of $\Phi_1$ should be
described by \eqref{eq:mono} {\it at all
times}. The quadrupole, on the other hand, is expected to deviate
significantly from the static prediction (\ref{eq:dipole})
when the black holes start moving fast.

As demonstrated in Fig.~\ref{fig:multipoles},
we find our results to be consistent with this picture.
Here we plot the monopole and quadrupole of
$\Phi_1$. The monopole part (left panel) captures the Coulomb field and
can thus be compared with the total charge of the system. It is constant
throughout the evolution to within numerical error and shows agreement
with the analytic prediction of Eq.~(\ref{eq:mono})
within numerical uncertainties;
we measure a slightly smaller value for the monopole field than expected from the total charge of the system,  but the measured value should increase with extraction radii and agree with the total charge expectation at infinity. This is consistent with the extrapolation of the measured value to infinity as shown in the figure. 
The quadrupole part (right panel) starts at a non-zero value in excellent
agreement with Eq.~(\ref{eq:dipole}), deviates substantially during the
highly dynamic plunge and merger stage and eventually rings down
towards the static limit $\phi_1^{20}=0$
as expected for a spherically symmetric charge distribution.
\begin{figure}[htbp]
\centering
\includegraphics[width=0.45\textwidth]{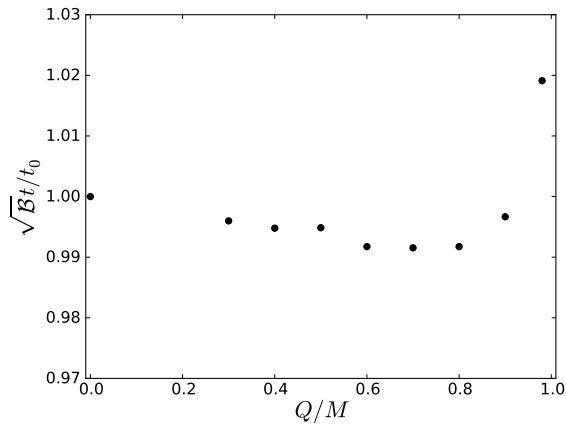}
\caption[]{Time for apparent horizon formation, re-scaled by the factor
  $\sqrt{{\cal B}}$ and the apparent horizon formation time $t_0$ for an
  electrically neutral binary.
  We note that the change in the quantity we plot is only, at most, of $2\%$. 
  The coordinate time itself, however, varies by a factor 5 as one goes from $Q=0$ to $Q=0.98M$.
  \label{fig:time}
}
\end{figure}

The analytic approximation of Sec.~\ref{classical_expectations} also
predicts a value for the time of collision (\ref{eq:timecollision})
for a given set of initial parameters. In particular, we see from this
prediction that for fixed initial separation $d$ and mass $M$ the collision
time scales with the charge as $t_{\rm collision} \sim 1/\sqrt{\mathcal{B}}$.
In comparing these predictions with our numerical results we face the
difficulty of not having an unambiguous definition of the separation
of the black holes in the fully general relativistic case. From
the entries in Table~\ref{tab:runs} we see that the proper distance $L$
varies only mildly for fixed coordinate distance $d$ up to $Q/M \approx 0.8$.
For nearly extremal values of $Q$, however, $L$ starts increasing
significantly as expected from our discussion at the start of this
section. We therefore expect the collision time of the numerical
simulations rescaled by $\sqrt{\mathcal{B}}/t_0$, where $t_0$
is the corresponding time for the uncharged case, to be close to unity
over a wide range of $Q/M$ and show some deviation close to $Q/M=1$.
This expectation is borne out in Fig.~\ref{fig:time} where we show
this rescaled collision time, determined numerically as the first
appearance of a common apparent horizon, as a function of $Q/M$.

\subsection{Waveforms: infall, merger and ringdown}

\begin{figure*}[htbp]
\centering
\includegraphics[width=0.45\textwidth]{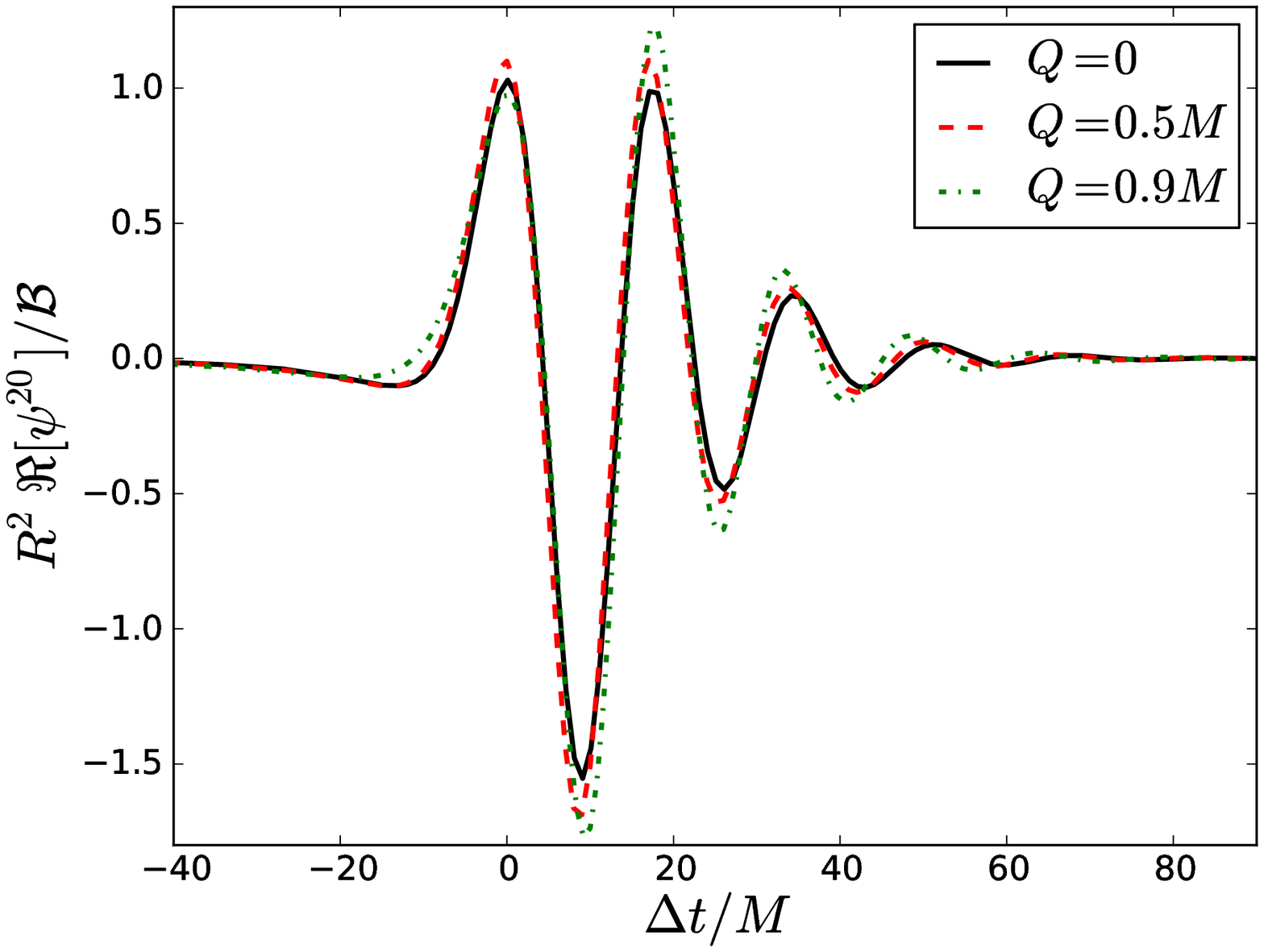}
\includegraphics[width=0.45\textwidth]{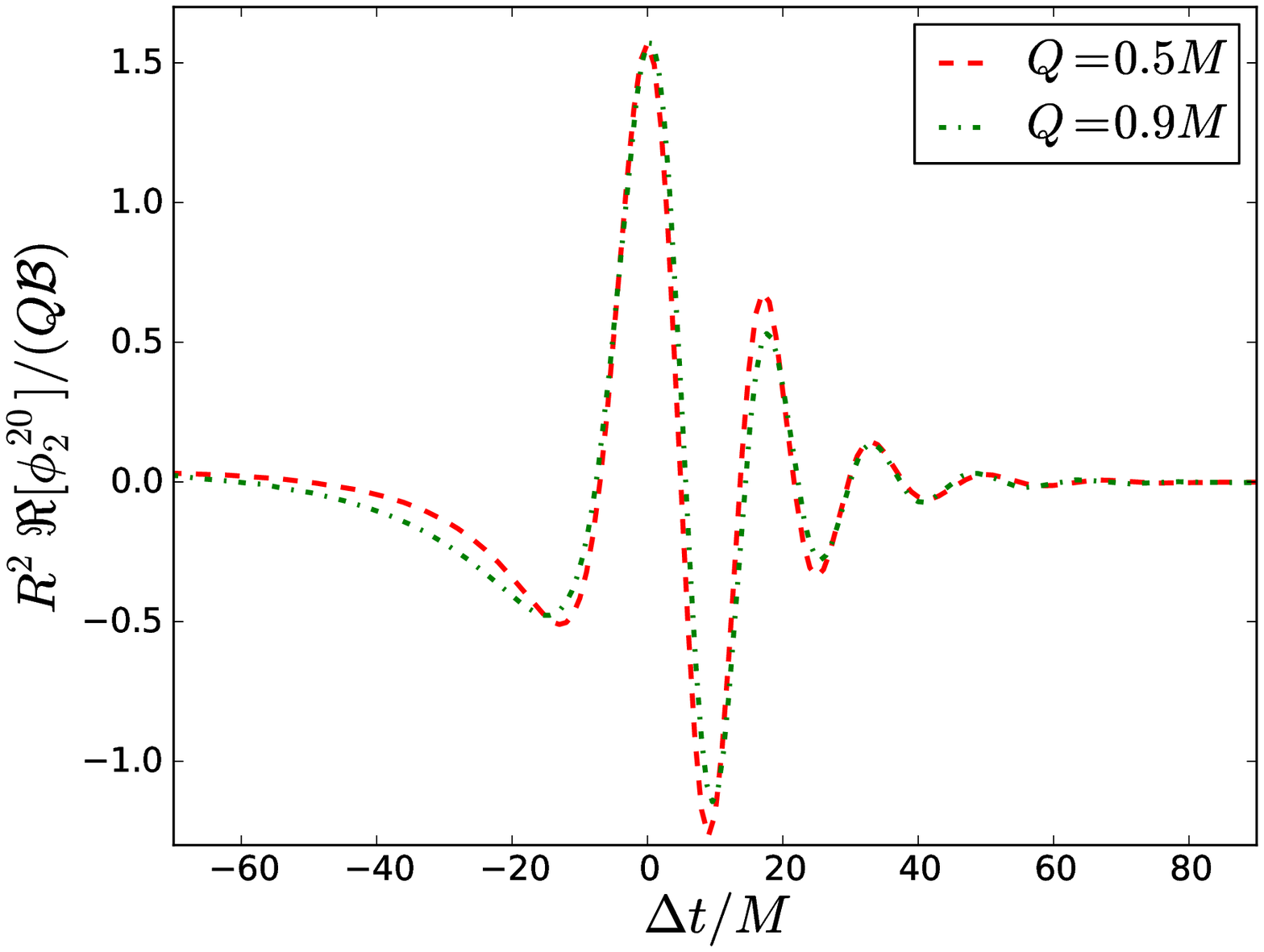}
\caption[]{Real part of the $(2,0)$ mode of $\Psi_4$ (left) and $\Phi_2$
  (right) extracted at $R_{\mathrm{ex}} = 100 M$.
  \label{fig:waveforms} }
\end{figure*}

The dynamical behaviour of all our simulations is qualitatively
well represented by the waveforms shown in Fig.~\ref{fig:waveforms}
for simulations d16q00, d16q05 and d16q09.
The panels show the real part of the
gravitational (left) and electromagnetic (right)
quadrupole extracted at $R_{\rm ex}=100~M$
as a function of time with $\Delta t=0$ defined as the time of
the global maximum of the waveform.
From the classical analysis \eqref{eq:power}, we expect the waveforms
$\Psi_4,\,\Phi_2$ to scale roughly with ${\cal B}$ and the mass
or charge of the black holes (the scaling with ${\cal B}$ is non-trivial, but both an analytic estimate and the numerical results indicate the scaling is approximately linear, which we shall therefore use for re-scaling the plots in the figure).
%

The early stages of the signals
are marked by the spurious radiation due to the construction of initial
data which we ignore in our analysis. Following a relatively weak phase
of wave emission during the infall of the holes, the radiation increases
\begin{table}[t]
  \centering
  \begin{tabular*}{0.45\textwidth}{@{\extracolsep{\fill}}ccc}
    \hline
    \hline
    $Q/M$ & $\omega_{1,2}^{\rm QNM}$ & $\omega_{1,2}^{\rm ext}$  \\
    \hline
    0       & $0.374 -0.0890i$    & $0.374 - 0.088 i $ \\
            & $0.458 -0.0950i$    &                    \\
    0.3     & $0.376 -0.0892i$    & $0.375-0.092i$     \\
            & $0.470 -0.0958i$    & $0.481-0.100i$     \\
    0.5     & $0.382 -0.0896i$    & $0.381-0.091i$     \\
            & $0.494 -0.0972i$    & $0.511-0.096i$     \\
    0.9     & $0.382 -0.0896i$    & $0.381-0.091i$     \\
            & $0.494 -0.0972i$    & ?  \\
    \hline
    \hline
  \end{tabular*}
  \caption{Comparison of the ringdown frequencies obtained from
    (i) perturbative calculations~\cite{Berti:2009kk} and
    (ii) fitting a two-mode profile to the numerically extracted waveforms.
    For $Q/M=0$ the electromagnetic modes are not excited. For values
    of $Q/M \ge 0.9$ the electromagnetic mode becomes so weak that we
    can no longer unambiguously identify it in the numerical data.}
  \label{tab:ringdown}
\end{table}
strongly during the black-hole merger around $\Delta t=0$ in the figure
and decays exponentially as the final hole rings down into a stationary
state. This overall structure of the signals is rather similar for the
electromagnetic and the gravitational parts and follows the main pattern
observed for gravitational-wave emission in head-on collisions of
uncharged black holes~\cite{Witek:2010xi,Witek:2010az}.

\begin{figure}[b]
\centering
\includegraphics[width=0.45\textwidth]{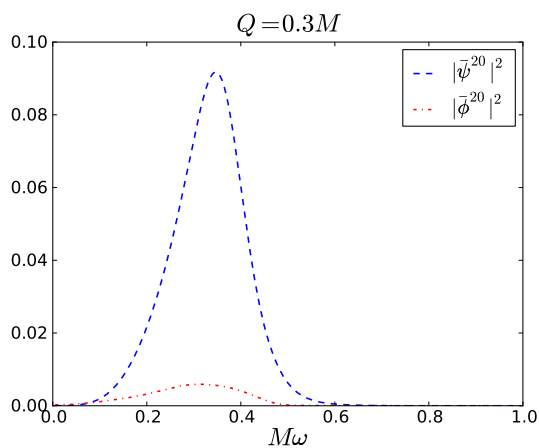}
\caption[]{Power spectrum for the gravitational (long dashed) and
electromagnetic (short dashed) quadrupole extracted from simulation~d08q03. 
Note that the spectrum peaks near the
fundamental ringdown frequency of the gravitational mode;
cf.~Table~\ref{tab:ringdown}.}
\label{fig:fourier}
\end{figure}

The final, exponentially damped ringdown phase is well described by
perturbation techniques~\cite{Berti:2009kk}. In particular, charged black
holes are expected to oscillate with two different types of modes, one
of gravitational and one of electromagnetic origin.
For the case of vanishing charge, the electromagnetic modes are
not present, but they generally couple for charged black holes,
and we expect both modes to be present
in the spectra of our gravitational and electromagnetic waveforms.
For verification we have fitted the late-stages of the waveforms to a
two-mode, exponentially damped sinusoid waveform
\begin{equation}
  f(t) = A_1 e^{-i\omega_1 t} + A_2 e^{-i\omega_2 t},
\end{equation}
where $A_i$ are real-valued amplitudes and $\omega_i$ complex
frequencies. The results are
summarized in Table~\ref{tab:ringdown} for selected values of the
charge-to-mass ratio of the post-merger black hole. Real and imaginary
parts of the fitted frequencies agree within a few percent or better
with the perturbative predictions. For the large value $Q/M$, however,
the wave signal is very weak and in such good agreement with a single
ringdown mode (the gravitational one) that we cannot clearly identify
a second, electromagnetic component.
This feature is explained once we understand how the total radiated
energy is distributed between the gravitational and the electromagnetic
channels. For this purpose, we plot in Fig.~\ref{fig:fourier}
the Fourier spectrum of the relevant wavefunctions or, more
precisely, their dominant quadrupole contributions obtained
for simulation~d08q03
$|\bar{\phi}^{20}|^2,|\bar{\psi}^{20}|^2$, where for any function $f$
\begin{equation}
  \bar{f}(\omega)=\int_{-\infty}^\infty e^{i\omega t}f(t)dt\,.
\end{equation}
It is clear from the figure that most of the energy is carried in
the fundamental gravitational-wave like mode with a peak
at approximately $\omega\sim 0.37$, close to the oscillation frequency
of the fundamental gravitational ringdown mode; see Table~\ref{tab:ringdown}.

\subsection{Radiated energy and fluxes}
\label{sec:fluxes}
%
\begin{figure}[b]
\centering
\includegraphics[width=0.45\textwidth]{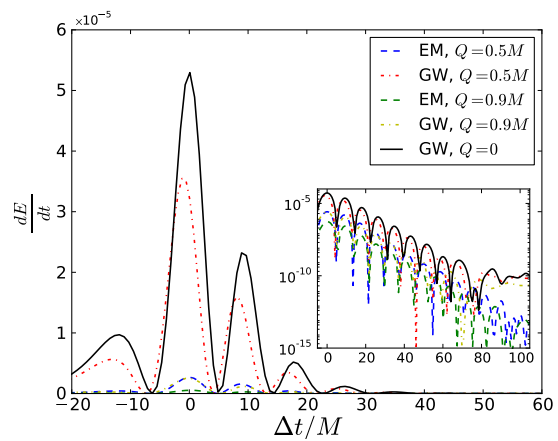}
\caption[]{Radiated fluxes for simulations d08q05, d08q09 and d08q00
  of Table~\ref{tab:runs}. We have aligned the curves in time such
  that their global maximum coincides with $t=0$. The inset shows the exact same plot with the $y$-axis in logarithmic units.
  \label{fig:Flux} }
\end{figure}
The  
electromagnetic and gravitational
wave fluxes are given by Eqs.~(\ref{eq:GW-flux}) and (\ref{eq:EM-flux}). We have
already noticed from the waveforms in Fig.~\ref{fig:waveforms} that
the electromagnetic signal follows a pattern quite similar to the
gravitational one. The same holds for the energy flux which is
shown in Fig.~\ref{fig:Flux} for a subset of our simulations
with $Q/M=0$, $0.5$ and $0.9$.
From the figure, as well as the numbers in Table~\ref{tab:runs},
we observe that the
energy carried by gravitational radiation decreases with increasing $Q/M$,
as the acceleration becomes smaller and quadrupole emission is suppressed, in
agreement with prediction \eqref{GWquadprediction}.

\begin{figure}[t]
\centering
\includegraphics[width=0.45\textwidth]{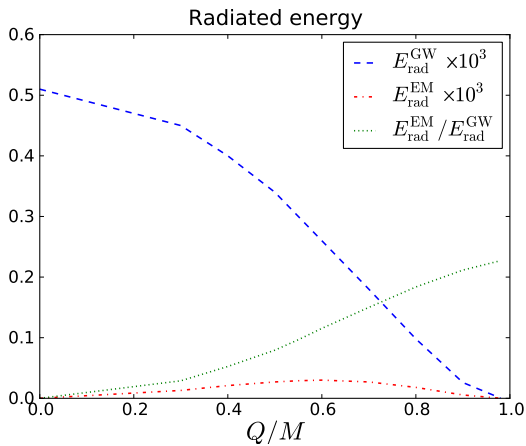}
\caption[]{Energy radiated in the gravitational and electromagnetic
  quadrupole as well as the ratio of the two as a function of $Q/M$.
  \label{fig:Energy_QM}}
\end{figure}

This is further illustrated in Fig.~\ref{fig:Energy_QM}, which illustrates the
radiated energy carried in the gravitational quadrupole and the
electromagnetic quadrupole as well as their ratio as functions of
the charge-to-mass ratio $Q/M$.
For the case of vanishing charge, the total radiated energy
is already known from the literature; e.g.~\cite{Witek:2010xi}.
The value increases mildly with the initial separation as a consequence
of the slightly larger collision velocity but is generally found to be
close to $E_{\rm rad}^{\rm GW}/M=0.055~\%$. Our values of $0.051~\%$
for $d/M\simeq 8$ and $0.055~\%$ for $d/M \simeq 16$ are in good
agreement with the literature. As we increase $Q/M$, however,
$E_{\rm rad}^{\rm GW}$
decreases significantly and for $Q/M=0.9$ ($0.98$) has dropped
by a factor of about $20$ ($10^3$) relative to the
uncharged case. For practical reasons, we have explored
the largest ratio $Q/M=0.98$ for the smaller initial separation
$d/M\simeq 8$ only; the near cancellation of the gravitational and
electromagnetic interaction and the resulting slow-down of the
collision lead to a very long infall stage with essentially zero
dynamics.

In contrast to the monotonically decreasing gravitational-wave energy,
the electromagnetic signal reaches a local maximum around $Q/M=0.6$,
an expected observation as the electromagnetic radiation necessarily
vanishes for $Q/M=0$ (no charge) and $Q/M=1$ (no acceleration)
but takes on non-zero values in the regime in between.
Closer analysis of our classical, flat-space calculation
\eqref{EMquadprediction} predicts a maximum electromagnetic radiation
output at
\begin{equation}
  Q_{\rm max}=\sqrt{\frac{\sqrt{329}-13}{14}}\,M\approx 0.605 M\,,
\end{equation}
in excellent agreement with the results of our NR simulations.

We finally consider the ratio of electromagnetic to
gravitational wave energy (dotted curve in Fig.~\ref{fig:Energy_QM}).
As predicted by our analytic calculation (\ref{prediction_ratio}),
this ratio increases monotonically with $Q/M$ for fixed separation $d$.
A fit of our numerical results yields
$E_{\rm rad}^{\rm EM}/E_{\rm rad}^{\rm GW}=0.27~Q^2/M^2$ and for
our largest value $Q/M=0.98$, we obtain a ratio of $0.227$
to be compared with $\sim 0.24$
as predicted by Eq.~(\ref{prediction_ratio}). Bearing in mind
the simplicity of our analytic model in Sec.~\ref{classical_expectations},
the quantitative agreement is remarkable.

\section{Final remarks}
\label{sec:conclusion}
We have performed a numerical study of collisions of charged black
holes with equal mass and charge in the framework of the fully
non-linear Einstein-Maxwell equations.
Our first observation is that the numerical relativity techniques
(formulation of the evolution equations,
gauge conditions and initial data construction)
developed for electrically neutral black-hole binaries can be straightforwardly
extended to successfully model charged binaries even for nearly extremal
charge-to-mass ratios $Q/M \lesssim 1$.
In particular, we notice the contrast with
the case of rotating black holes with nearly extremal
spin which represents a more delicate task for state-of-the-art
numerical relativity; cf.~Refs.~\cite{Lovelace:2011nu,Lousto:2012es} for the
latest developments on this front.
This absence of difficulties for charged holes is not entirely unexpected.
Considering the construction of initial data, for instance,
an important difference arises in the customary choice of conformally flat
Bowen-York initial data~\cite{Bowen:1980yu}
which greatly simplifies the initial data problem. While the
Kerr solution for a single rotating black hole does not
admit conformally flat slices~\cite{Garat:2000pn} and therefore
inevitably results in spurious radiation, especially for large
spin parameters, this difficulty does not arise for charged, but
non-rotating black holes; cf.~Eq.~(\ref{eq:conformalfactor})
and~\cite{Graves:1960zz}.

The excellent agreement between the classical calculation for the energy
emission and the numerical results reported here, allow for an investigation of
Cosmic Censorship close to extremality. If we take two black holes with $M_1=M_2=M/2$,
$Q_1=Q_2=(M-\delta)/2$
and we let them fall from infinity, to first order in $\delta$ we get
\begin{equation}
\begin{aligned}
Q_{\rm tot}& = M - \delta \\
M_{\rm tot}& = M - E_{\rm rad}
\end{aligned} \,.
\end{equation}
Now, the classical result~\eqref{GWquadprediction} implies that the dominant
term for the radiated energy is $E_{\rm rad} \sim \mathcal{B}^{5/2} M \sim
(\delta/M)^{5/2} M$.
Thus we get
\begin{equation}
\frac{Q_{\rm tot}}{M_{\rm tot}} \simeq 1 
                                  - \frac{\delta}{M} 
                                  + k \left(\frac{\delta}{M}\right)^{5/2}\,,
\end{equation}
where $k$ is a constant.  We get the striking conclusion that Cosmic Censorship
is preserved for charged collisions of nearly extremal holes ($\delta \ll M$),
on account of the much longer collision time, which yields much lower velocities
and therefore much lower energy output. The differences between the cases of
spinning mergers and charged collisions are interesting. In the former case,
naked singularities are avoided by radiation carrying away more angular momentum
(via orbital hangup~\cite{Campanelli:2006uy}).  In the latter case, our results
suggest that naked singularities are avoided by the smaller radiation emission,
due to the smaller accelerations involved in the infall.

It is even possible to construct binary initial data in closed analytic
form, analogous to that of Brill-Lindquist data, for the special
case of non-spinning binaries with equal charge-to-mass ratio
starting from rest and we have restricted our present study to this case.
Specifically, we have evolved a sequence of binaries with $Q/M$
varying from zero to values close to extremality. Starting with
the electrically neutral case, where our gravitational wave emission
$E_{\rm rad}^{\rm GW}/M=0.055~\%$ agrees well with the literature,
we observe a monotonic decrease of the emitted gravitational wave
energy as we increase $Q/M$. For our largest value $Q/M=0.98$,
$E_{\rm rad}^{\rm GW}$ is reduced by about three orders of magnitude,
as the near cancellation of the gravitational and electromagnetic
forces substantially slows down the collision. In contrast, the
radiated electromagnetic energy reaches a maximum near $Q/M=0.6$ but
always remains significantly below its gravitational counterpart.
Indeed, the ratio $E_{\rm rad}^{\rm EM}/E_{\rm rad}^{\rm GW}$
increases monotonically with $Q/M$ and approaches about $25~\%$
in the limit $Q/M \rightarrow 1$. We find all these results to be
in remarkably good qualitative {\em and} quantitative agreement
with analytic approximations obtained in the framework of the
dynamics of two point charges in a Minkowski background. This
approximation also predicts that the collision time relative
to that of the uncharged case scales $\sim \sqrt{1-Q^2/M^2}$
which is confirmed within a few percent by our numerical simulations.

Our present study paves the way for various future extensions.
Quite naturally, it will be important to consider more generic
types of initial data in order to tackle some of the issues discussed in the
Introduction. A non-zero boost, for instance, will allow us to
study both binary black hole systems that will coalesce into a Kerr-Newman
black hole and the impact of electric charge on the dynamics of
wave emission (electromagnetic and gravitational)
in high energy collisions. In this context the robustness of our
simulations is particularly encouraging, as we have not encountered
stability issues as observed in the study of black-hole collisions
in higher-dimensional spacetimes~\cite{Zilhao:2011yc}.

A further interesting extension presently under study is the case
of oppositely charged black holes. Quite likely, the remarkable
accuracy of our simple analytic models is in part due to the
relatively small, ``non-relativistic'' collision speeds caused by
the electric repulsion of the equal charges.
Furthermore, the gravitational quadrupole formula (\ref{GWquadprediction})
will still apply for opposite charges, but
then ${\cal B}=1+Q^2/M^2$, and the formula predicts an enhancement of
almost two orders of magnitude in the gravitational radiation emitted
when going from $Q=0$ to $Q=M$ (without accounting for
additional contributions due to dipole radiation and to ``Bremsstrahlung'' by
accelerated charges). This would release about $3\%$ of the total
center of mass energy as gravitational waves. Even more impressive is the
possibility of observing a huge splash of electromagnetic energy
when both holes are nearly extremal. The area theorem, which yields a poor estimate in the neutral case, bounds the total radiation to be less than $\sim 65\%$ the CM energy; how close one gets to this number is up to nonlinear evolutions of the kind reported in this work.

\begin{acknowledgments}
We are indebted to Leonardo Gualtieri, Steve Giddings and Emanuele Berti for fruitful
discussions on this topic.
We further thank the anonymous referee for suggesting the calculation of the
charge-to-mass ratio of the merged hole, in particular in the near
extremal limit.
M.Z.\ would like to thank the hospitality of
the Perimeter Institute for Theoretical Physics, through the Visiting
Graduate Fellows program, where this work was done.  The authors thank
the Yukawa Institute for Theoretical Physics at Kyoto University,
where parts of this work were completed during the YITP-T-11-08 on
``Recent advances in numerical
and analytical methods for black hole dynamics''.  M.Z.\ is funded by
FCT through grant SFRH/BD/43558/2008.  U.S.\ acknowledges support from
the Ram\'on y Cajal  Programme and Grant FIS2011-30145-C03-03  of the
Ministry of Education and Science of Spain,
the NSF TeraGrid and XSEDE Grant No.~PHY-090003,
RES Grant Nos.~AECT-2012-1-0008 and AECT-2011-3-0007 through the
Barcelona Supercomputing Center and CESGA Grant Nos.~ICTS-200 and ICTS-221.
This work was supported
by the {\it DyBHo--256667} ERC Starting Grant, the {\it CBHEO--293412}
FP7-PEOPLE-2011-CIG Grant,  the {\it NRHEP--295189} FP7-PEOPLE-2011-IRSES
Grant, and by  FCT -- Portugal through projects PTDC/FIS/098025/2008,
PTDC/FIS/098032/2008, PTDC/FIS/116625/2010 and CERN/FP/116341/2010,
the Sherman Fairchild Foundation to Caltech as well as NSERC through a
Discover Grant and CIFAR. Research at Perimeter Institute is supported
through Industry Canada and by the Province of Ontario through the
Ministry of Research \& Innovation. Computations were performed on the
Blafis cluster at Aveiro University, the Milipeia in Coimbra, the
Lage cluster at Centro de F\'\i sica do Porto, the Bifi Cluster of
the University of Zaragoza, the CESGA Cluster Finis Terrae, NICS Kraken
and the SDSC Cluster Trestles.

\end{acknowledgments}

\appendix

\section{Geodesic slicing}
\label{sec:geo}

In the usual Schwarzschild-like coordinates, the Reissner-Nordstr\"om line element and electromagnetic potential are given by

\begin{equation}
\begin{aligned}
  ds^2 & =-f(R)dt^2+\frac{dR^2}{f(R)}+R^2d\Omega_2 \, , \\
  A & = -\frac{Q}{R} dt \,,
\end{aligned}  \label{eq:RN-schw}
\end{equation}
where $f(R) = 1- \frac{2M}{R} + \frac{Q^2}{R^2} $.
For a radially in-falling massive particle, starting from rest at $R=R_0$, the
energy per unit mass is $\sqrt{f(R_0)}$. The geodesic equation (for in-falling particles) may then be written as
\begin{equation}
  \frac{dt}{d\tau}=\frac{\sqrt{f(R_0)}}{f(R)} \, ,
      \qquad \frac{dR}{d\tau} = - \sqrt{f(R_0)-f(R)} \, .
      \label{eq:geodesics}
\end{equation}
With these equations and the initial condition $R(\tau = 0) = R_0$, we can numerically integrate this system and thus have $R=R(\tau, R_0)$.
Assuming such a coordinate transformation, $(t,R) \to (\tau,R_0)$, the metric takes the form
\begin{equation}
  ds^2 = -d\tau^2 + \left( \frac{\partial R(\tau,R_0) }{\partial R_0}
  \right)^2 \frac{dR_0^2}{f(R_{0})} + R(\tau,R_0)^2 d\Omega_2 \, .
\end{equation}
It remains now to perform the coordinate transformation $R_0 \to r$ that guarantees the metric an isotropic form at $\tau = 0$. 
This can be accomplished with
\begin{equation}
\frac{dr}{r} = \frac{dR_0}{R_0 \sqrt{f(R_0)}} \,;
\end{equation}
integrating we obtain
\begin{equation}
\begin{aligned}
R_0(r) & = r \left[
  \left(1+\frac{M}{2r}\right)^2 - \frac{Q^2}{4r^2}
\right]
\,.
\end{aligned}
\end{equation}
The final form for the metric is then
\begin{widetext}
  \begin{equation}
    ds^2 = -d\tau^2 + \left( \frac{R_0(r)}{r} \right)^2
    \left[
      \left(\frac{\partial R(\tau,R_0) }{\partial R_0} \right)^2 dr^2
      + \left( \frac{r}{R_0(r)} \right)^2 R(\tau, R_0(r))^2 d\Omega_{2}
    \right] \,.
  \end{equation}
Since, by assumption, $R(\tau=0) = R_0$, $\left.\frac{\partial R}{\partial R_0} \right|_{\tau=0} = 1 $, this metric is indeed in isotropic form at $\tau = 0$.
\end{widetext}

\bibliographystyle{myutphys}
\bibliography{num-rel}

\end{document}